\documentstyle[aps,multicol,psfig]{revtex}
\begin{document}
\draft
\title{Velocity distribution in granular gases of viscoelastic particles}

\author{ Nikolai V. Brilliantov$^{1,2}$ and  Thorsten P\"oschel$^{1}$}
\address{$^1$Humboldt-Universit\"at zu Berlin, Institut f\"ur Physik,
Invalidenstra{\ss}e 110, \\ D-10115 Berlin, Germany}
\address{$^2$Moscow State University, Physics Department, 
Moscow 119899, Russia}
\date{\today}
\maketitle
\begin{abstract}
The velocity distribution in a homogeneously cooling granular gas has been studied in the viscoelastic regime when the restitution coefficient of colliding particles depends on the impact velocity.  We show that for viscoelastic particles the simple scaling hypothesis is violated, i.e., that the time dependence of the velocity distribution does not scale with the mean square velocity as in the case of particles interacting via a constant restitution coefficient. The deviation from the Maxwellian distribution does not depend on time monotonously. For the case of 
small dissipation we detected two regimes of evolution of the velocity distribution function:
Starting from the initial Maxwellian distribution, the deviation first increases with time 
on a collision time-scale saturating  at some maximal value; then it 
decays to zero on much larger time-scale  which corresponds to the temperature relaxation. 
For larger values of the dissipation parameter there appears an additional intermediate relaxation 
regime.  Analytical calculations for 
small dissipation agrees well with the results of a numerical analysis.

\end{abstract}
\begin{multicols}{2}
\section{Introduction}

The statistical properties of granular gases have been intensively studied in recent time, in particular with respect to cluster formation process, e.g. \cite{Cluster} and other structure formation, e.g. \cite{ErnstEtAl}. In the present paper we are concerned with dynamical processes in granular gases which precede clustering, i.e. in the homogeneously cooling state  (HCS). In difference to the state when particles form clusters and other long range structures, in the HCS (due to its definition) one may drop the explicit spatial dependence of the statistical properties, which simplifies an application of standard methods of the gas  kinetic gas theory. Granular gases in the HCS were intensively investigated recently (see e.g. \cite{Proceed} for a review) focusing on the velocity distribution function which is one of the most important characteristics of the system of granular particles.  It has been argued that the distribution function might deviate from the Maxwellian
~\cite{GoldshteinShapiro95,EsipovPoeschel:97}, and this deviation has been also 
quantified ~\cite{GoldshteinShapiro95,NoijeErnst:97,BrilliantovPoeschel:99:a2}. 

In all of these studies a constant restitution coefficient, characterizing the energy loss due to a particle collision was assumed. The restitution coefficient relates the velocities of the colliding particles before a collision $\vec{v}_1$, $\vec{v}_2$ to the velocities after the collision $\vec{v}_1^*$, $\vec{v}_2^*$:
\begin{eqnarray}
\label{directcoll}
\label{v1v2v1*v2*}
&&\vec{v}_1^*=\vec{v}_1-
\frac12 (1+\epsilon) (\vec{v}_{12} \cdot \vec{e})\vec{e} \nonumber \\
&&\vec{v}_2^*=\vec{v}_2+
\frac12 (1+\epsilon) (\vec{v}_{12} \cdot \vec{e})\vec{e} 
\end{eqnarray}
where $\vec{v}_{12}=\vec{v}_1-\vec{v}_2$ is the relative velocity and the 
unit vector $\vec{e} = \vec{r}_{12}/ \left|\vec{r}_{12}\right|$ gives the direction of the inter-center vector $\vec{r}_{12}=\vec{r}_{1}-\vec{r}_{2}$ at the 
instant of the collision. Strictly speaking the restitution coefficient 
$\epsilon$ as introduced in Eq.~(\ref{directcoll}) describes the collision of 
{\em smooth} inelastic particles, when only the normal component 
$(\vec{v}_{12} \cdot \vec{e})$ of the relative velocity $\vec{v}_{12}$ changes. 
Therefore, it is termed as {\em normal} restitution coefficient. Using  
the tangential restitution coefficient 
\cite{CundalStrack79,BSHP,LudingZippelius:98}, one can 
account for the change of tangential component of the relative velocity at 
the collision of rough inelastic particles. In what follows  we assume that the particles are smooth and the dynamics of a collision is completely described by the change of the normal component of the relative velocity. 

Experiments, as well as theoretical studies show, however, that 
$\epsilon$ noticeably depends on the impact velocity $\vec{v}_{12}$ 
\cite{epsviavel,C10,Zipp99}; 
even a dimension analysis shows that the assumption 
of the constant restitution coefficient contradicts physical reality 
\cite{Taguchi:1992JDP,Rozaetal}. This dependence may cause rather important 
consequences for various problems in granular gas dynamics 
\cite{generalGG,BrilliantovPoeschel:1998d}. The problem of the restitution coefficient's 
dependence on the impact velocity has been addressed in \cite{BSHP,BSHP1},
where the generalization of the Hertz contact problem was developed 
for the collision of viscoelastic particles (scaling analysis of this 
dependence has been also addressed in \cite{C10}). The generalized 
Hertz collision equation derived in \cite{BSHP} has been solved 
analytically to obtain the velocity-dependent restitution coefficient
\cite{TomThor} 
\begin{equation}
\label{epsC1C2}
\epsilon=1-C_1 A \alpha^{2/5} \left|\vec{v}_{12} \cdot \vec{e}\right|^{1/5}
+C_2A^2 \alpha^{4/5}\left|\vec{v}_{12} \cdot \vec{e}\right|^{2/5}  \mp \cdots
\end{equation}
with 
\begin{equation}
\alpha = \left( \frac32 \right)^{3/2} \, 
\frac{Y\sqrt{R^{\rm eff}}}{m^{\rm eff} (1-\nu^2)}
\end{equation}
where $Y$ is the Young modulus, $\nu$ is the Poisson ratio, 
$R^{\rm eff}=R_1R_2/(R_1+R_2)$, $m^{\rm eff}=m_1m_2/(m_1+m_2)$
($R_{1/2}$ and $m_{1/2}$ are radii and masses of colliding particles), 
$A$ is the dissipative constant, which depends on the material parameters
(see \cite{BSHP} for details). Numerical values for the constants $C_1$ and 
$C_2$ obtained in \cite{TomThor} may be also written in a more convenient 
form \cite{Rozaetal}:
\begin{eqnarray}
\label{C1C2}
C_1 &=& \frac{ \Gamma(3/5)\sqrt{\pi}}{2^{1/5}5^{2/5} \Gamma(21/10)} =
1.15344,\\ 
\label{C1C22}
C_2&=&\frac{3}{5}C_1^2\, .
\end{eqnarray}

Although the next-order coefficients $C_3=0.315119C_1^{3}$, 
$C_4=0.161170C_1^4$, are now available \cite{Rozaetal}, we assume that 
the dissipative constant  $A$ is small enough to ignore these high-order 
terms.   

The aim of the present study is to analyze how the impact-velocity dependent 
restitution coefficient given by Eq.~(\ref{epsC1C2}) for the collision of viscoelastic spheres, influences the velocity distribution in a granular gas of identical particles in HCS. To address this problem we use the Sonine polynomials 
expansion for the velocity distribution function and analyze the 
time-dependence of the expansion coefficients. 

In Sec.~\ref{sec:Sonine} we introduce the necessary variables, briefly sketch the method of Sonine polynomial expansion and summarize the knowledge about the velocity distribution function in granular gases under the assumption of a constant restitution coefficient. In Sec. \ref{sec:Kinetic} we analyze the Boltzmann equation for the granular gas with the velocity- dependent $\epsilon$ in the HCS and calculate the first few coefficients of the Sonine polynomials expansion. We show that these coefficients occur to be time-dependent, so that the velocity distribution function does not have a simple scaling form. In Sec. \ref{sec:Evolution} we consider  the time evolution of temperature and of the velocity distribution.  The high-velocity tail of the distribution function is 
analyzed in Sec. \ref{sec:Tail}. In Conclusion we summarize our findings.  Some technical detail of the calculations are given in the Appendixes. 

\section{Sonine polynomial expansion for granular gases}
\label{sec:Sonine}

For granular gases where the particles interact via a restitution coefficient 
$\epsilon={\rm const} $ it was 
argued that the velocity distribution $f(\vec{v}, t)$ has the scaling form, 
i.e. that its time-dependence may be written as (here we follow notations
of Ref.\cite{NoijeErnst:97})
\begin{equation}
\label{veldis}
f(\vec{v}, t)=\frac{n}{v_0^d(t)} \tilde{f} \left( \frac{v}{v_0(t)}\right) 
\end{equation}
where $n$ is the number density of the granular gas, $v_0(t)$ is the 
thermal velocity, defined in terms of the temperature of the granular gas
\begin{equation}
\label{deftemp}
T(t)=\frac12 mv_0^2(t)
\end{equation}
$m$ is the mass of the granular particles, and $d$ is the dimension. The 
temperature is related to the second moment of the velocity distribution 
in the same way as for equilibrium molecular systems:
\begin{equation}
\label{deftemp1}
\frac{d}{2} n T(t)=\int d \vec{v} \frac{m v^2}{2} f(\vec{v},t)\,.
\end{equation}

Then the expansion of the scaling function $\tilde{f}(\vec{c})$ (where 
$\vec{c} \equiv \vec{v}/v_0(t)$) in terms of the Sonine polynomials 
reads \cite{GoldshteinShapiro95,NoijeErnst:97}
\begin{equation}
\label{Soninexp}
\tilde{f}(\vec{c})=\phi(c) \left\{1 + \sum_{p=1}^{\infty} a_p S_p(c^2) \right\}\,,
\end{equation}
where $\phi(c) \equiv \pi^{-d/2} \exp(-c^2)$ is the Maxwellian distribution 
for the rescaled velocity. The Sonine polynomials $S_p(c^2)$ satisfy the 
orthogonality conditions
\begin{equation}
\label{Soninortog}
\int d \vec{c} \phi (c) S_p(c^2)S_{p^{\prime}}(c^2) 
= \delta_{pp^{\prime}}{\cal N}_p
\end{equation}
with $\delta_{pp^{\prime}}$ being the Kronecker delta and with the 
normalization constant ${\cal N}_p$ \cite{GoldshteinShapiro95,NoijeErnst:97}. 
For dimension $d=3$ which is addressed here the first few Sonine polynomials
read
\begin{eqnarray}
\label{Soninfewfirst}
&&S_0(x)=1 \nonumber \\
&&S_1(x)=-x^2 +\frac32 \\
&&S_2(x)=\frac{x^2}{2}-\frac{5x}{2}+\frac{15}{8}
\end{eqnarray}
The coefficients $a_p$ of the expansion may be found as the polynomial 
moments of the function $\tilde{f}(\vec{c})$ 
\cite{GoldshteinShapiro95,NoijeErnst:97}:
\begin{equation}
\label{Sonincoef}
a_p=\frac{1}{{\cal N}_p} \int d \vec{c} S_p(c^2) \tilde{f}(\vec{c})
\end{equation}
The coefficients $a_p$ do not depend on time for a {\em constant} 
restitution coefficient \cite{remarktime}. These were first applied for the granular 
gas in  Ref.\cite{GoldshteinShapiro95} and then recalculated recently \cite{NoijeErnst:97}: 

\begin{eqnarray}
\label{Sonina1a2}
a_1&=&0 \\
a_2&=&\frac{16(1-\epsilon)(1-2\epsilon^2)}{9+24d +8\epsilon d +41\epsilon+30(1-\epsilon)\epsilon^2 }\,.
\label{Sonina1a2a}
\end{eqnarray}
The first relation (\ref{Sonina1a2}) follows from the definition of 
the temperature of the granular gas (this we explain in more detail below), while 
Eq. (\ref{Sonina1a2a}) has been obtained 
within the linear approximation with respect to $a_2$. Complete analysis, which goes beyond 
the linear approximation, has been performed \cite{BrilliantovPoeschel:99:a2}, and it 
has been shown \cite{BrilliantovPoeschel:99:a2} that the linear solution (\ref{Sonina1a2a}) 
is rather accurate for the whole range of $\epsilon$ with a maximal deviation from a total 
one less than $10\%$ \cite{remarkroots}. All the higher-order coefficients were neglected 
under assumption of small deviations from the Maxwellian distribution. Since 
$a_p $ do not depend on time, the  scaling form of the velocity distribution function (\ref{veldis}) 
persists  with time for the case of $\epsilon={\rm const}$.

Since the average velocity of a granular gas decreases due to permanently 
decreasing temperature, the ``typical'' restitution coefficient will increase 
with time as it follows from Eq. (\ref{epsC1C2}). Thus one can expect that the 
coefficients of the Sonine polynomials expansion, 
which depend on the restitution coefficient (see e.g. (\ref{Sonina1a2a})) 
should also change with time. 
This conclusion, however, contradicts the assumption, that the scaling function 
(\ref{Soninexp}) does not depend on time and implies that the 
common scheme of calculation of the Sonine polynomials expansion 
coefficients breaks  down if $\epsilon$ is not a constant. For for the latter 
case, one needs to develop a more general approach. 

\section{Kinetic equation for the coefficients of the Sonine polynomials expansion}
\label{sec:Kinetic}
We start from the Enskog-Boltzmann equation for the distribution 
function $f(\vec{r},\vec{v},t)$ for a granular gas of inelastic spheres which in the force-free case does not depend on $\vec{r}$, Hence, one can write 
\cite{NoijeErnst:97,resibua}
\begin{eqnarray}
\label{collint} 
\frac{\partial}{\partial t}f(\vec{v}_1,t) &=&g_2(\sigma)\sigma^2 \int d \vec{v}_2 \int d\vec{e} \, 
\Theta(-\vec{v}_{12} \cdot \vec{e}) \left|\vec{v}_{12} \cdot \vec{e}\right| \times\nonumber\\
&&\times \left\{\chi f(\vec{v}_1^{**},t)f(\vec{v}_2^{**},t)-
f(\vec{v}_1,t)f(\vec{v}_2,t) \right\} \nonumber \\
& \equiv& g_2(\sigma)I(f,f)
\end{eqnarray}
where $\sigma$ is the diameter of the particles. 
The contact value of the two-particle correlation function, $g_2(\sigma)=(2-\eta)/2(1-\eta)^3$ \cite{CarnahanStarling} (with $\eta=\frac16\, \pi n \sigma^3$ being the packing fraction) accounts for the increasing 
collision frequency due to the excluded volume effects. $\Theta(x)$ is the Heaviside step-function. The velocities $\vec{v}_1^{**}$ and 
$\vec{v}_2^{**}$ refer to the pre-collisional velocities of the so-called 
inverse collision, which results with $\vec{v}_1$ and $\vec{v}_2$ as the 
after-collisional velocities (the relation between these velocities are 
similar to that of Eq. (\ref{directcoll}), but with the impact-velocity 
dependent restitution coefficient, see Appendix A). Finally the factor 
\begin{eqnarray}
\label{CHI}
\chi &=& 1 + \frac{11}{5}C_1 A \alpha^{2/5} \left|\vec{v}_{12} \cdot \vec{e}\right|^{1/5}\nonumber \\
&+&\frac{66}{25}C_1^2 A^2 \alpha^{4/5} \left|\vec{v}_{12} \cdot \vec{e}\right|^{2/5} 
+\cdots
\end{eqnarray}
in the gain term appears respectively from the Jacobian of the transformation 
$d\vec{v}_1^{**}d\vec{v}_2^{**} \to d\vec{v}_1 d\vec{v}_2$ and from 
the relation between the lengths of the collisional cylinders 
$\epsilon \left|\vec{v}_{12}^{**} \cdot \vec{e}\right| dt=\left|\vec{v}_{12} \cdot \vec{e}\right|dt$ 
(see Appendix A for details). For the constant restitution coefficient 
$\chi = 1/\epsilon^2 ={\rm const}$

Some important properties of the collisional integral hold also for the 
case of the impact-velocity dependent restitution coefficient. Namely, it may 
be shown that the relation
\end{multicols}
\begin{eqnarray}
\label{deraver} 
&&\frac{d}{dt} \left< \psi(t) \right> =\int d\vec{v}_1 \psi (\vec{v}_1)
\frac{\partial}{\partial t} f(\vec{v}_1, t) = 
\int d\vec{v}_1 \psi (\vec{v}_1) I(f,f)\\
&&=\frac{g_2(\sigma)\sigma^2}{2} \int d\vec{v}_1d\vec{v}_2 \int d\vec{e}
\Theta(-\vec{v}_{12} \cdot \vec{e}) \left|\vec{v}_{12} \cdot \vec{e}\right|
f(\vec{v}_1, t)f(\vec{v}_2, t) \Delta \left[ \psi( \vec{v}_1)+
\psi( \vec{v}_2) \right] \nonumber 
\end{eqnarray}
\begin{multicols}{2}
\noindent holds true, where 
$\left< \psi(t) \right> \equiv \int d\vec{v}_1 \psi (\vec{v}) f(\vec{v}, t)$ is 
the average of some function $\psi (\vec{v})$, and 
$\Delta \psi(\vec{v}_i) \equiv \left[\psi(\vec{v}_i^*)-\psi(\vec{v}_i) \right]$
denotes  change of $\psi( \vec{v}_i)$ in a direct collision.

Now we analyze the scaling ansatz (\ref{veldis}) for the velocity distribution 
function. Using this ansatz and performing calculations similar to that 
in Ref.\cite{NoijeErnst:97}, one would find corresponding expressions for the 
coefficients $a_p$ of the Sonine polynomial expansion. These would occur to be 
time-dependent due to permanently decreasing average velocity of the cooling gas
and thus permanently increasing effective value of the restitution coefficient. 
This however means that the 
simple scaling (\ref{veldis}) for the velocity distribution function 
does not hold for the case 
of interest. Technically, as we show below, this follows from the additional 
time-dependence of the factor $\chi$ in the collisional integral, which does not 
depend on time for $\epsilon={\rm const}$.

Thus, it seems natural to write the three dimensional distribution function in the general form 
\begin{equation}
\label{genscal}
f(\vec{v}, t)=\frac{n}{v_0^3(t)}\tilde{f}(\vec{c}, t)
\end{equation}
with 
\begin{equation}
\label{genSoninexp}
\tilde{f}(\vec{c}, t)
=\phi(c) \left\{1 + \sum_{p=1}^{\infty} a_p(t) S_p(c^2) \right\}
\end{equation}
and find then equations for the {\em time-dependent} coefficients $a_p(t)$. 

Substituting (\ref{genscal}) into the Boltzmann equation (\ref{collint})
we obtain
\begin{eqnarray}
&&\frac{1}{v_0^2} \frac{dv_0}{dt}
\left(3 + c_1 \frac{\partial}{\partial c_1} \right) \tilde{f}(\vec{c}_1, t) +
\frac{1}{v_0} \frac{\partial}{\partial t} \tilde{f}(\vec{c}_1, t) =\nonumber\\
&=&g_2(\sigma) \sigma^2 n \tilde{I}\left( \tilde{f}, \tilde{f} \right)\,,
\label{Boltzred}
\end{eqnarray}
where we define the dimensionless collisional integral
\begin{eqnarray}
\tilde{I}\left( \tilde{f}, \tilde{f} \right)&=&
\int d \vec{c}_2 \int d\vec{e} \, 
\Theta(-\vec{c}_{12} \cdot \vec{e}) \left|\vec{c}_{12} \cdot \vec{e}\right|\times\nonumber\\
&\times&\left\{\tilde{\chi} \tilde{f}(\vec{c}_1^{**},t) \tilde{f}(\vec{c}_2^{**},t)-
\tilde{f}(\vec{c}_1,t)\tilde{f}(\vec{c}_2,t) \right\} \,.
\label{dimlcolint}
\end{eqnarray}
The reduced factor $\tilde{\chi}$
\begin{equation}
\label{chiscal}
\tilde{\chi} = 1 + 
\frac{11}{5}C_1 \delta^{\prime} \left|\vec{c}_{12} \cdot \vec{e}\right|^{1/5}
+\frac{66}{25}C_1^2 \delta^{\prime \, 2} \left|\vec{c}_{12} \cdot \vec{e}\right|^{2/5}
+\cdots
\end{equation}
depends now on time via a quantity
\begin{equation}
\label{deltaprime}
\delta^{\, \prime} (t) \equiv A \alpha^{2/5} \left[2T(t)\right]^{1/10} 
\equiv \delta \left[2T(t)/T_0 \right]^{1/10} \,.
\end{equation}
Here $\delta \equiv A \alpha^{2/5}\left(T_0\right)^{1/10}$, $T_0$ is the initial 
temperature, and for simplicity we assume the particles to be of unit mass, $m=1$. 

The rate of change of the thermal velocity $dv_0/dt$ in (\ref{Boltzred}) 
may be expressed in terms of the temperature decay rate $dT/dt$, which reads 
according to the definitions (\ref{deftemp},\ref{deftemp1}) and  relation 
(\ref{deraver}) for the time derivatives of averages
\begin{equation}
\label{dTdt}
\frac{dT}{dt}=\frac13g_2(\sigma) \sigma^2 n v_0^3
\int d \vec{c}_1 c_1^2 
\tilde{I}\left( \tilde{f}, \tilde{f} \right) =-\frac23 BT\mu_2\,.
\end{equation} 
We define here $B=B(t) \equiv v_0(t) g_2(\sigma) \sigma^2 n$ and introduce the 
moments of the dimensionless collision integral: 
\begin{equation}
\label{mup}
\mu_p \equiv - \int d \vec{c}_1 c_1^p 
\tilde{I}\left( \tilde{f}, \tilde{f} \right)\ .
\end{equation}
With this notations we recast (\ref{Boltzred}) into the form:
\begin{equation}
\label{geneqveldis}
\frac{\mu_2}{3} 
\left(3 + c_1 \frac{\partial}{\partial c_1} \right) \tilde{f}(\vec{c}, t) +
B^{-1} \frac{\partial}{\partial t} \tilde{f}(\vec{c}, t) =
\tilde{I}\left( \tilde{f}, \tilde{f} \right)\,.
\end{equation}
Note that contrary to the case of $\epsilon = {\rm const}$, where 
$\chi=1/\epsilon^2={\rm const}$, the factor $\chi$  depends now on time,  
which  does not allow to write the collision integral in terms of only 
scaling variables. This implies time dependence of all the moments $\mu_p$ 
(which were time-independent for constant restitution coefficient) and 
correspondingly causes  time dependence of the Sonine polynonomials expansion
coefficients $a_p$.  

Multiplying both sides of Eq. (\ref{geneqveldis}) with $c_1^p$ and integrating 
over $ \vec{c}_1$ we obtain:
\begin{equation}
\label{momeq}
\frac{\mu_2}{3} p \left< c^p \right> -B^{-1}\sum_{k=1}^{\infty} 
\dot{a}_k \nu_{kp} = \mu_p\,,
\end{equation}
where integration by parts has been performed and where we define
\begin{equation}
\label{nukp}
\nu_{kp} \equiv \int \phi(c) c^p S_k(c^2) d\vec{c}
\end{equation}
and 
\begin{equation}
\label{cpdef}
\left< c^p \right>  \equiv \int c^p \tilde{f}(\vec{c}, t)  d\vec{c}\,.
\end{equation}
The calculation of the $\nu_{kp}$ is straightforward; the first few of them  read:
\begin{equation}
\label{nu2224}
\nu_{22}=0; \qquad \nu_{24}=\frac{15}{4}\,.
\end{equation}
The odd moments $\left< c^{2n+1} \right> $ are zero, while the even ones, 
$\left< c^{2n} \right> $, may be expressed in terms of $a_k$ with 
$0 \leq k  \leq n$. This follows from the fact that $c^{2n}$ may be written 
as a sum of Sonine polynomials $S_k(c^2)$ with $0 \leq k  \leq n$  and from the 
orthogonality condition (\ref{Soninortog}). Namely, using
$c^2=\frac32 S_0(c^2)-S_1(c^2)$ together with the expansion 
(\ref{genSoninexp}) and condition (\ref{Soninortog}), one easily finds
\begin{eqnarray}
\left< c^2 \right> &=& \int d \vec{c} \phi (c) \left[\frac32 S_0(c^2)-S_1(c^2) \right]
\left\{ \sum_{k=0}^{\infty} a_k S_k(c^2) \right\}\nonumber\\
 &=&\frac32 - \frac32 a_1 
\label{c2}
\end{eqnarray}
with $a_0 =1$ and where we use the normalization constant ${\cal N}_1=\frac32$ [see Eq. (\ref{Soninortog})].
From the definitions of temperature and of the thermal velocity (\ref{deftemp}), (\ref{deftemp1})
follows that $\left< c^2 \right> = \frac32$ 
(see also (\ref{cpdef})). Then Eq. (\ref{c2}) implies $a_1=0$ in 
accordance with Ref.\cite{NoijeErnst:97}. Similar considerations yield
\begin{equation}
\label{c4}
\left< c^4 \right> = \frac{15}{4}\left( 1 + a_2 \right)\,.
\end{equation} 
The moments $\mu_p$ may be also expressed in terms of coefficients 
$a_2, a_3, \cdots$, therefore, the system 
(\ref{momeq}) is an infinite (but closed) set of equations for these 
coefficients. 

It is not possible to get a general solution to the problem. However, since
the dissipative parameter $\delta$ is supposed to be small, the deviations from the 
Maxwellian distribution are not presumably large. Thus we assume, that one can neglect all the 
high-order terms in the expansion (\ref{genSoninexp}) with $p>2$. Then (\ref{momeq}) 
is an equation for the coefficient $a_2$. For $p=2$ Eq. (\ref{momeq}) converts into identity
since $\left< c^2 \right>=\frac32$, $a_1=0$ and due to (\ref{nu2224}). For $p=4$ we obtain
\begin{equation}
\label{eqa2}
\dot{a}_2-\frac43\, B\mu_2 \left(1+a_2 \right)+\frac{4}{15}B\mu_4 =0\,,
\end{equation} 
where the relations (\ref{nu2224}) and (\ref{c4}) have been used. In Eq. (\ref{eqa2}) $B$ 
depends on time as 
\begin{equation}
\label{BT}
B(t)=(8 \pi)^{-1/2} \tau_c(0)^{-1}[T(t)/T_0]^{1/2}\,,
\end{equation} 
where $T_0$ is the initial temperature and $\tau_c(0)$ is the initial mean-collision time
\begin{equation}
\label{tauc0}
\tau_c(0)^{-1}=4 \pi^{1/2}g_2(\sigma) \sigma^2 n T_0^{1/2}\, .
\end{equation} 
The time evolution of the temperature is determined by Eq. (\ref{dTdt}), i.e., by the 
time dependence of $\mu_2$. 

The time-dependent coefficients $\mu_p(t)$ may be expressed in terms of $a_2$ according to 
definition (\ref{mup}) and the approximation $\tilde{f}= \phi (c) \left[ 1+a_2(t)S_2\left(c^2\right)\right]$. One obtains:
\begin{eqnarray}
\label{mupa2}
\mu_p&=&-\frac12 \int d\vec{c}_1\int d\vec{c}_2 \int d\vec{e} 
\Theta(-\vec{c}_{12} \cdot \vec{e}) \left|\vec{c}_{12} \cdot \vec{e}\right| \phi(c_1) \phi(c_2) 
\nonumber \\
&&\times\left\{1+a_2\left[S_2(c_1^2)+S_2(c_2^2) \right] + a_2^2\,S_2(c_1^2)S_2(c_2^2) \right\}\times\nonumber\\
&&\times\Delta (c_1^p+c_2^p) 
\end{eqnarray}
with the definition of $\Delta (c_1^p+c_2^p)$ given above. After long and tedious 
calculations (details are given in Appendix B) one arrives at the following result for the 
moments:

\begin{equation}
\label{MU2A}
\mu_2=\delta^{\, \prime } \left[{\cal A}_1 + {\cal A}_2 a_2+ {\cal A}_3 a_2^2 \right] -
\delta^{\, \prime \,2 } \left[{\cal A}_4 + {\cal A}_5 a_2+ {\cal A}_6 a_2^2 \right] 
\end{equation} 
and 

\begin{eqnarray}
\label{MU4A}
\mu_4=\left[{\cal B}_1 + {\cal B}_2 a_2+ {\cal B}_3 a_2^2 \right] 
+\delta^{\, \prime } \left[{\cal B}_4 + {\cal B}_5 a_2+ {\cal B}_6 a_2^2 \right] \\
-\delta^{\, \prime \,2 } \left[{\cal B}_7 + {\cal B}_8 a_2+ {\cal B}_9 a_2^2 \right]
\nonumber 
\end{eqnarray}
where ${\cal A}_n$ and ${\cal B}_n$ are pure numbers. The coefficients ${\cal A}_n$ read
\begin{equation}
\label{A1A6}
\begin{tabular}{lll}
${\cal A}_1=\omega_0$ & $\quad {\cal A}_2=\frac{6}{25}\omega_0$ & $\quad {\cal A}_3=\frac{21}{2500} \omega_0$ \\[0.3cm]
${\cal A}_4=\omega_1$ & $\quad  {\cal A}_5=\frac{119}{400}\omega_1$ & $\quad {\cal A}_6=\frac{4641}{640000}\omega_1$
\end{tabular}
\end{equation}
with 
\begin{eqnarray}
\omega_0 &\equiv& 2 \sqrt{2 \pi} 2^{1/10} \Gamma \left (\frac{21}{10} \right)C_1=6.48562 \ldots\\
\omega_1 &\equiv& \sqrt{2 \pi} 2^{1/5} \Gamma \left (\frac{16}{5} \right)C_1^2=9.28569 \ldots, 
\end{eqnarray}
and the coefficients ${\cal B}_n$ are
\begin{equation}
\label{B1B3}
\begin{tabular}{lll}
${\cal B}_1=0$ & $\quad {\cal B}_2=4\sqrt{2 \pi}$ &$\quad {\cal B}_3=\frac{1}{8}\sqrt{2 \pi}$  \\[0.3cm]
${\cal B}_4= \frac{56}{10}\omega_0$ & $\quad  {\cal B}_5=\frac{1806}{250}\omega_0$ & $\quad {\cal B}_6=\frac{567}{12500}\omega_0$ \\[0.3cm]
${\cal B}_7= \frac{77}{10}\omega_1$ & $\quad  {\cal B}_8=\frac{149054}{13750}\omega_1$ & $\quad {\cal B}_9=\frac{348424}{5500000}\omega_1$
\end{tabular}
\end{equation}

Thus, Eqs. (\ref{eqa2}) and (\ref{dTdt}), together with 
Eqs. (\ref{deltaprime}), (\ref{BT}), (\ref{MU2A}) and  
(\ref{MU4A}) form a  closed set to find the time evolution of the temperature and 
coefficient $a_2$. We want to stress an important difference for the time evolution 
of temperature for the case of the impact-velocity dependent restitution coefficient, 
compared to that of the constant restitution coefficient. 
In the former case it is coupled to the time evolution of the coefficient $a_2$, while  
in the latter case there is no such coupling since $a_2={\rm const}$. 
This coupling may lead in some case to rather peculiar time-dependence of the temperature. 
The problem of the time dependence of temperature and the velocity distribution function will be discussed in detail in the following section.

\section{Time evolution of temperature and of the velocity distribution function}
\label{sec:Evolution}

To analyze the time evolution of the temperature and of the coefficient $a_2$, 
characterizing the velocity distribution function, we introduce the reduced 
temperature $u(t) \equiv T(t)/T_0$ and recast the set (\ref{eqa2}), (\ref{dTdt}) into 
the form
\begin{eqnarray}
\label{genseteq1}
&&\dot{u}+\tau_0^{-1}u^{8/5}\left( \frac53 +\frac25 a_2+
\frac{7}{500}a_2^2 \right) -\nonumber\\
&-&\tau_0^{-1}q_1 \delta \, u^{17/10}
\left( \frac53 +\frac{119}{240}a_2 +\frac{1547}{128000}a_2^2 \right) =0\\
\label{genseteq2}
&&\dot{a}_2-r_0u^{1/2}\mu_2 \left(1 +a_2 \right) + \frac15 r_0u^{1/2}\mu_4=0\,.
\end{eqnarray}
The characteristic time  
\begin{equation}
\label{tau0}
\tau_0^{-1}=\frac{16}{5} q_0 \delta \cdot \tau_c(0)^{-1}
\end{equation} 
describes the time evolution of the temperature (see below), with 
\begin{eqnarray}
q_0&=&2^{1/5}\Gamma(21/10)C_1/8=5^{-2/5}\sqrt{\pi}\Gamma(3/5)/8\nonumber\\
&=&0.173318\ldots\\
r_0 &\equiv& \frac{2}{3 \sqrt{2 \pi}} \tau_c(0)^{-1}\\
q_1 &\equiv& 2^{1/10} (\omega_1/\omega_0) =1.53445 \ldots
\end{eqnarray}
To obtain these
equations we use the 
expressions for $\mu_2(t)$, $B(t)$ and for coefficients ${\cal A}_n$. Note that the 
characteristic time $\tau_0$ is $\delta^{-1} \gg 1$ times larger than the mean collision 
time $\tau_c(0)$. 

We will find the  solution to these equations as expansions in terms of the small 
dissipative parameter $\delta$ (see Eq.(\ref{deltaprime})):

\begin{eqnarray}
\label{expudel}
&&u=u_0+ \delta \cdot u_1 +\delta^2 \cdot u_2 +\cdots \\
\label{expadel}
&&a_2=a_{20}+\delta \cdot a_{21}+\delta^2 \cdot a_{22} +\cdots
\end{eqnarray} 
Substituting Eqs. (\ref{expudel}), (\ref{expadel}), (\ref{MU2A}) and (\ref{MU4A}) into 
Eqs. (\ref{genseteq1}), (\ref{genseteq2}), one can solve these equations perturbatively, 
for each order of $\delta$. Collecting terms of the order of ${\cal O}(1)$ we obtain:

\begin{eqnarray}
\label{udel1}
&&\dot{u}_0 +\tau_0^{-1} 
\left(\frac53 +\frac25 a_{20}+\frac{7}{500}a_{20}^2 \right)u_0^{8/5}=0 \\
\label{adel1}
&&\dot{a}_{20}+r_1u_0^{1/2}\left( a_{20} +\frac{1}{32}a_{20}^2 \right)=0
\end{eqnarray} 
where  
\begin{equation}
r_1 \equiv \frac15 r_0 {\cal B}_2=\frac{8}{15}\tau_c(0)^{-1}
\end{equation}
and we use the 
definition of $r_0$, and expressions (\ref{B1B3}) for ${\cal B}_2$ and 
${\cal B}_3$, which are zero-order coefficients in the expansion of $\mu_4$ on 
$\delta$. Changing variables 
\begin{equation}
t \to \tau =\int_0^{t}dt^{\prime}u_0^{1/2}(t^{\prime})
\end{equation}
in Eq. (\ref{adel1}) one finds the solution 
of  this (Riccati) equation:
\begin{equation}
\label{a20gen}
a_{20}(t)=
\frac{a_{20}(0)}{\left[1+\frac{1}{32}a_{20}(0) \right]e^{\tau}-\frac{1}{32}a_{20}(0)}
\end{equation} 
According to Eq. (\ref{udel1}) the characteristic time scale for $u_0(t)$ is $\tau_0 \gg \tau_c(0)$, therefore, for $t \sim \tau_c(0) \ll \tau_0$ one can approximate $u(t)=T(t)/T_0 \approx 1$. Moreover, if the initial deviation 
from the Maxwellian distribution is not large, i.e. $a_{20}(0)/32\ll 1$, one can 
approximate for this time interval:
\begin{equation}
\label{a20tsmall}
a_{20}(t) \approx a_{20}(0)e^{-4t/5 \tau_{\rm E}(0)}\,, 
\end{equation} 
with $\tau_{\rm E} =\frac32 \tau_c$ being the Enskog relaxation time. 
Therefore, $a_{20}(t)$ vanishes for $t \sim \tau_0  \gg \tau_c(0) $. This refers to the relaxation of an initially non-Maxwellian velocity distribution to the Maxwellian one. Note that the relaxation occurs within few collisions per particle, similarly to the relaxation of molecular gases. 

We now assume that the initial distribution is Maxwellian, i.e., that $a_{20}(0)=0$ for
$t=0$. Then the deviation from the Maxwellian distribution originates from the 
inelasticity of the particle collisions. For the case $a_{20}(0)=0$ 
(and thus $a_2(t)=0$, see Eq.(\ref{a20gen})) 
 the solution to Eq. (\ref{udel1}) reads
\begin{equation}
\label{T(t)del1}
u_0(t) = \frac{T(t)}{T_0}= \left(1 + \frac{t}{\tau_0} \right)^{-5/3}
\end{equation} 
which coincides with the time-dependence of the temperature obtained previously 
using scaling arguments \cite{TomThor} (up to a constant $\tau_0$ which may 
not be determined by scaling arguments). 

For the order ${\cal O} (\delta)$ we obtain:
\begin{eqnarray}
\dot{u}_1 +\frac8{3 \tau_0} u_0^{3/5}u_1+\frac2{5\tau_0} u_0^{8/5}a_{21}
-\frac5{3\tau_0}q_1 u_0^{17/10} =0 \label{udel2} \\
\label{adel2}
\dot{a}_{21}+r_1u_0^{1/2} a_{21} + r_2 u_0^{3/5} =0
\end{eqnarray} 
with 
\begin{equation}
r_2 \equiv \left(\frac{4}{15}\right)2^{1/10}(8 \pi)^{-1/2} \left( {\cal B}_4-5{\cal A}_1\right) \tau_c^{-1}(0).\end{equation}

For $t \ll \tau_0$ we have $u_0 \approx 1$ and Eq. (\ref{adel2}) reduces to 
\begin{equation}
\label{a21tsmall}
\dot{a}_{21}+r_1a_{21}=-r_2
\end{equation} 
with the solution:
\begin{equation}
\label{a21tsmallsol}
a_{21}(t)=-\frac{r_2}{r_1} \left( 1- e^{-r_1t} \right)=
-h \left( 1- e^{-4t/5\tau_{\rm E}(0)} \right)
\end{equation} 
where 
\begin{equation}
h \equiv r_2/r_1 =\left(\frac{3}{10}\right)\Gamma\left(\frac{21}{10}\right)2^{1/5}C_1=0.415964
\end{equation}
and we used the
definitions of $r_1$, $r_2$ and the values of ${\cal A}_k$ and ${\cal B}_k$ 
given above. As it follows from Eq. (\ref{a21tsmallsol}), after a transient 
time of the order of few collisions per particle, i.e. for 
 $\tau_{\rm E}(0) < t \ll \tau_0$, $a_{2}(t)$ saturates to the value 
$a_{2}=-h \delta =-0.415964 \delta $,
i.e. it changes only slowly on the time-scale $\sim \tau_c(0)$.

For $t \gg \tau_0$ the rescaled temperature 
varies $u_0 \approx (t/\tau_0)^{-5/3}$ [see Eq. (\ref{T(t)del1})], 
and Eq. (\ref{adel2}) reads
\begin{equation}
\label{a21tlarge}
\dot{a}_{21}+r_1\left(t/\tau_0 \right)^{-5/6}a_{21}=
-r_2\left(t/\tau_0 \right)^{-1}\,.
\end{equation} 
Using the power-law ansatz 
\begin{equation}
a_{21}(t) \sim \left(t/\tau_0 \right)^{-\nu}
\end{equation}
the asymptotic analysis of Eq. (\ref{a21tlarge}) yields the 
exponent $\nu =1/6$ and an estimate for the prefactor. Thus, we find for 
$t  \gg \tau_0$:
\begin{equation}
\label{a21tlargesol}
a_{21}(t)=-\frac{r_2}{r_1} \left( t/\tau_0  \right)^{-1/6}=
-h \left( t/\tau_0  \right)^{-1/6}
\end{equation} 
Therefore,
$a_{21}(t)$ decays to zero on the
time-scale $\sim \tau_0$, i.e., slowly on the time-scale 
$\sim \tau_c(0) \ll \tau_0$. The velocity distribution, thus, tends asymptotically 
to the Maxwellian distribution. 

One can also find the general solution of Eq. (\ref{adel2}):
\begin{eqnarray}
a_{21}(t)&=&-6\tau_0r_2 \exp \left\{ -6 \tau_0 r_1 \left(1+t/\tau_0 \right)^{1/6} \right\}\times\nonumber\\
&\times& \int^{6\tau_0r_1(1+t/\tau_0)^{1/6}}_{6\tau_0r_1} \frac{e^{x}}{x} \,  dx\,.
\label{a21gensol}
\end{eqnarray}
Noticing that 
\begin{eqnarray}
6 \tau_0 r_1&=& (q_0 \delta)^{-1}\\ 
6 \tau_0 r_2&=& \frac{12}{5} \delta^{-1}
\end{eqnarray}
due to the definitions or $r_1$, $r_2$ and $\tau_0$, 
one can write for $a_2(t)=\delta \cdot a_{21}(t)$ in linear with respect to 
$\delta$ approximation:
\begin{equation}
\label{a21gensolLi}
a_2(t)=-\frac{12}{5}w(t)^{-1}
\left\{ {\rm Li} \left[ w(t) \right]-{\rm Li} \left[ w(0) \right] \right\}
\end{equation} 
where 

\begin{equation}
\label{w(t)}
w(t) \equiv \exp \left[ \left(q_0 \delta \right)^{-1} \left(1+t/\tau_0 \right)^{1/6} \right]
\end{equation} 
and ${\rm Li}(x)$ is the logarithmic integral. 
It is not difficult to show that from the general expression 
(\ref{a21gensolLi}) both 
limiting dependencies (\ref{a21tsmallsol}) for $t \ll \tau_0$ and 
(\ref{a21tlargesol}) for $t \gg \tau_0$ are reproduced.

We could not find the general solution for $u_1(t)$, however, one can obtain the solution for $t \gg \tau_0$. Substituting asymptotic expressions
$u_0(t) \simeq (t/\tau_0)^{-5/3}$ and $a_{21}(t) \simeq -h(t/\tau_0)^{-1/6}$ into 
Eq. (\ref{udel2}) for $u_1(t)$ we recast this equation into the form
\begin{equation}
\label{u1asymp}
\dot{u}_1+\frac83 (t/\tau_0)^{-1}u_1=
\left(\frac25 h +\frac53 q_1 \right) (t/\tau_0)^{-17/6}
\end{equation}
Again a power-law ansatz $u_1(t) \sim (t/\tau_0)^{\alpha}$ allows to obtain 
both, the exponent $\alpha =11/6$ as well as the corresponding prefactor. The result 
for $u_1(t)$ for $t \gg \tau_0$ reads
\begin{equation}
\label{u1asympsol}
u_1(t)=\left( \frac{12}{25}h+2q_1 \right)(t/\tau_0)^{-11/6}=
3.26856(t/\tau_0)^{-11/6}
\end{equation}
where we used the above results for the constants $h$ and $q_1$. From the last 
equation one can see  how the coupling between the temperature
and the 
velocity distribution influences the evolution of temperature. Indeed, if there 
were  no such coupling, there would be no coupling term in Eq. (\ref{udel2}), and 
thus, no contribution from $\frac{12}{25}h$ to the prefactor of $u_1(t)$ in 
(\ref{u1asympsol}). This would noticeably change  the time behavior of $u_1(t)$. 
On the other hand, the leading term in the time dependence of temperature, $u_0(t)$, is not affected by this kind of coupling.

On Fig.~\ref{fig:a2} and Fig.~\ref{fig:T} we show the time dependence of the coefficient
$a_2(t)$ of the Sonine polynomial expansion and of the temperature of the
granular gas. The analytical findings are compared with the numerical
solution of the system (\ref{genseteq1},\ref{genseteq2}). As it follows  from the
figures the analytical theory reproduces fairly well the
numerical results for the case of small $\delta$.

As it follows from Fig.~\ref{fig:a2},
for small $\delta$ the following scenario of evolution of the velocity
distribution takes place for a force-free granular gas.
The initial Maxwellian distribution evolves to a non-Maxwellian distribution, with
the discrepancy between these two characterized by the second coefficient of the
Sonine polynomials expansion $a_2$. The deviation from the Maxwellian distribution
(described by $a_2$) quickly grows, until it saturates after a few collisions per
particle at a ``steady-state'' value. At this instant the deviation from the Maxwellian
distribution is maximal, with the value $a_2 \approx - 0.4 \delta$ (Fig.~\ref{fig:a2}, top).
This refers to the first ``fast'' stage of the evolution, which takes place on a
mean-collision time-scale $\sim \tau_c(0)$. After this maximal deviation is reached,
the second ``slow'' stage of the evolution starts. At this stage $a_2$ decays to zero on
``slow'' time scale $\tau_0 \sim \delta^{-1} \tau_c(0) \gg \tau_c(0)$, which
corresponds to the time  scale of the temperature evolution (Fig.~\ref{fig:a2}, middle); the decay of the
coefficient $a_2(t)$ in this regime occurs according to a power law $ \sim t^{-1/6}$
(Fig.~\ref{fig:a2}, bottom). Asymptotically the
Maxwellian distribution would be achieved, if the clustering process did not occur.

Fig.~\ref{fig:T} illustrates the significance of the first-order 
correction $u_1(t)$ in the time-evolution
of temperature. This becomes more important as the dissipation parameter $\delta$
grows (Fig.~\ref{fig:T} top, Fig.~\ref{fig:T}  middle). At large times the results of the first-order theory
(with $u_1(t)$ included) practically coincide with the numerical results, while
zero-order theory (without $u_1(t)$) demonstrates noticeable 
deviations (Fig.~\ref{fig:T} bottom). 
\begin{minipage}{8cm}
\begin{figure}[htbp]
\centerline{\psfig{figure=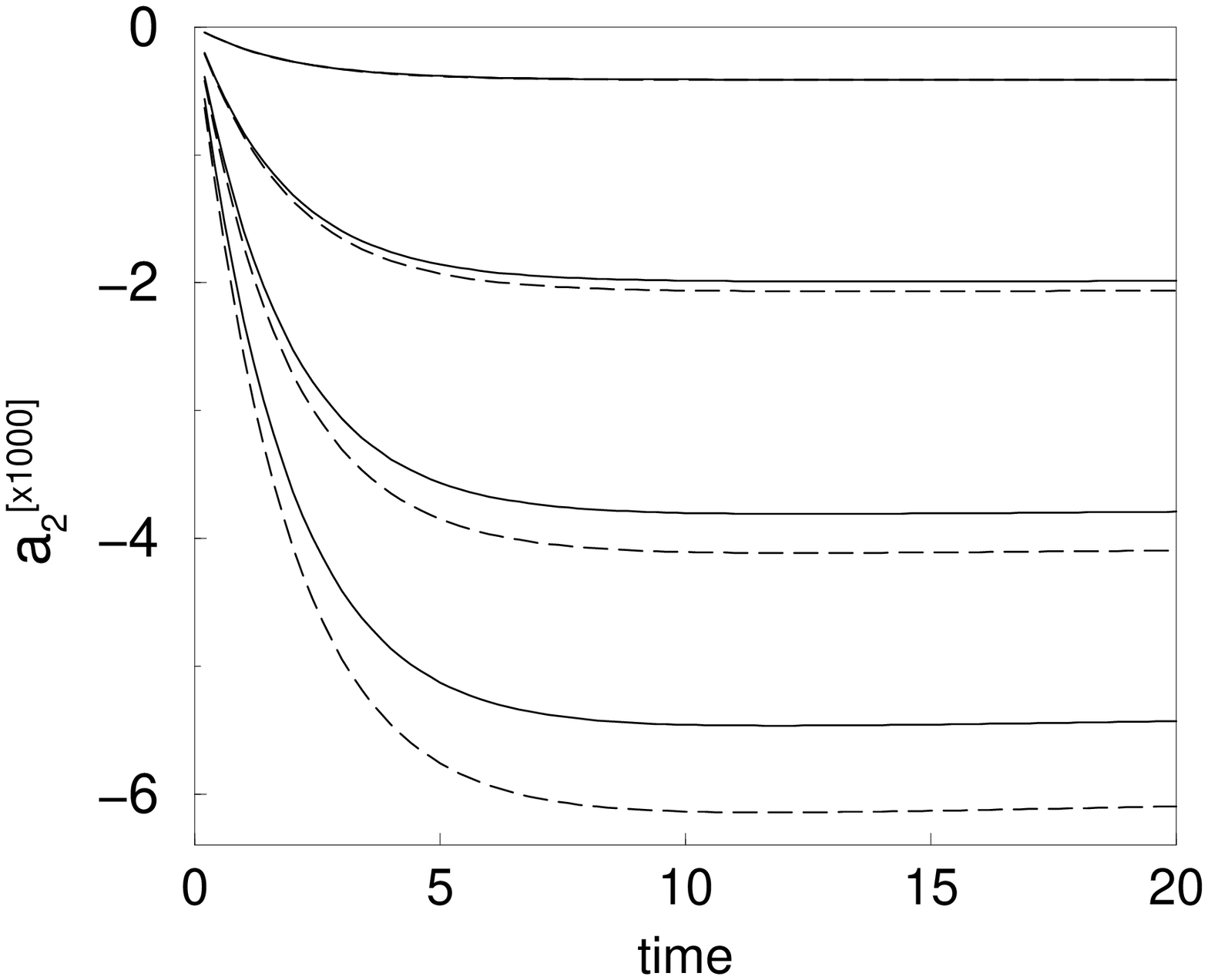,width=7.5cm}}
\centerline{\psfig{figure=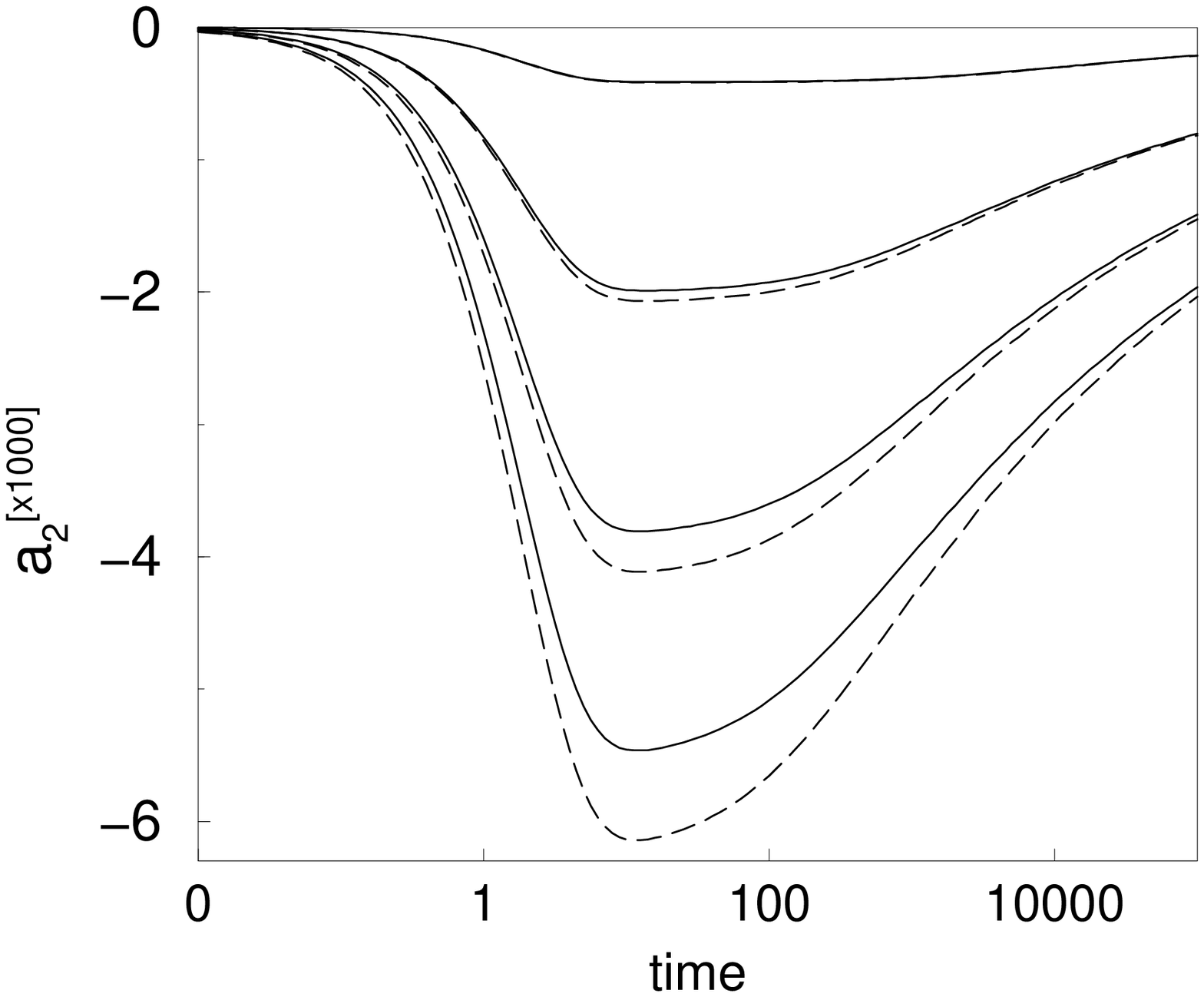,width=7.5cm}}
\centerline{\psfig{figure=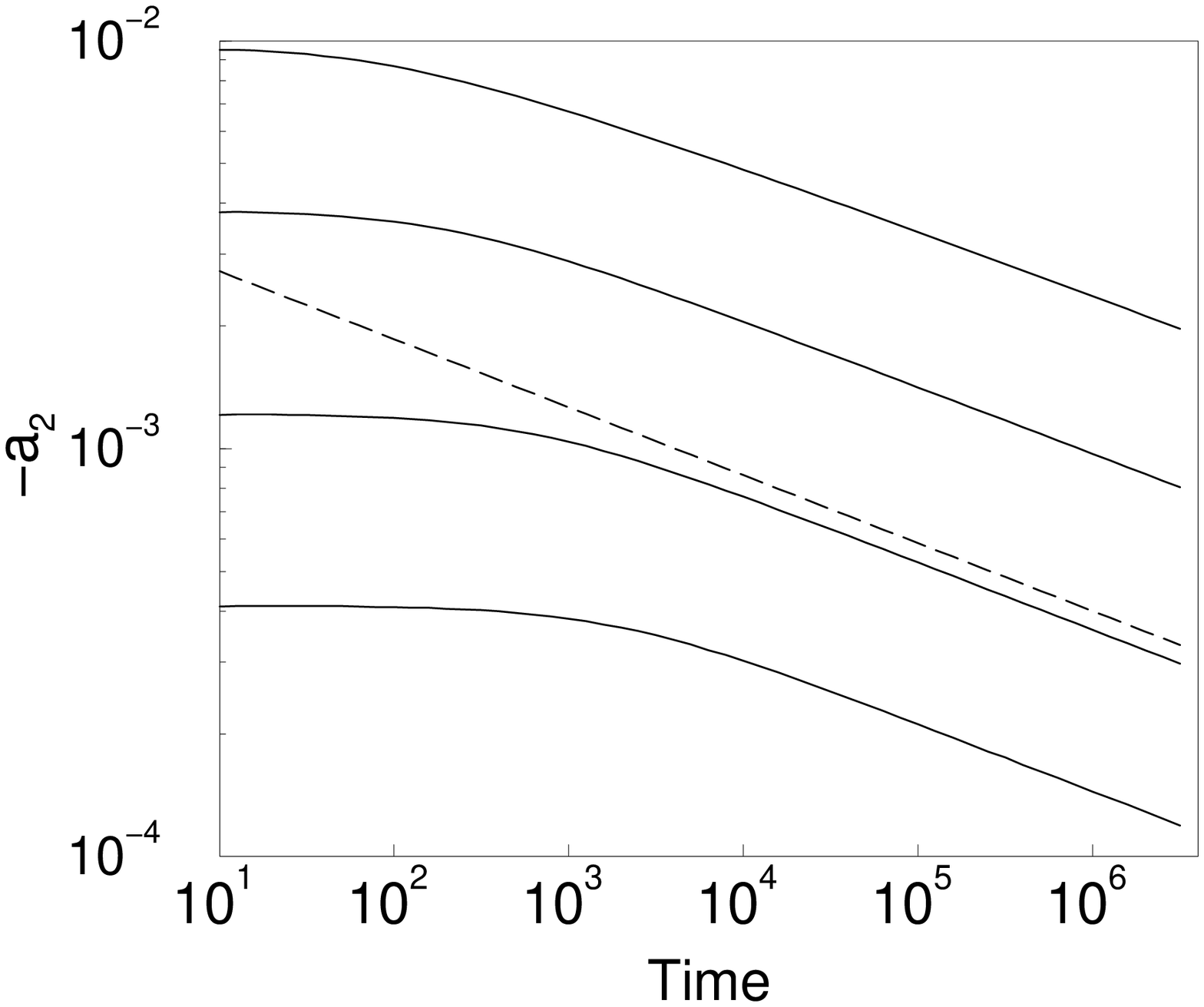,width=7.5cm}}
  \caption{Time dependence of the second coefficient of the Sonine polynomial
expansion $a_2(t)$. Time is given in units of mean collisional time
$\tau_c(0)$. (Top): $a_2 \times 1000$ (solid lines) for
$\delta =0.001, 0.005, 0.01, 0.015$
(top to bottom) together with the linear approximation (dashed lines); (Middle): the same as (Top) 
but for larger times; (Bottom): $-a_2(t)$ 
over time (log-scale) for $\delta =0.03, 0.01, 0.003, 0.001$ (top to
bottom) together with the power-law asymptotics $\sim t^{-1/6}$.}
  \label{fig:a2}
\end{figure}
\end{minipage}

\begin{minipage}{8cm}
\begin{figure}[htbp]
\centerline{\psfig{figure=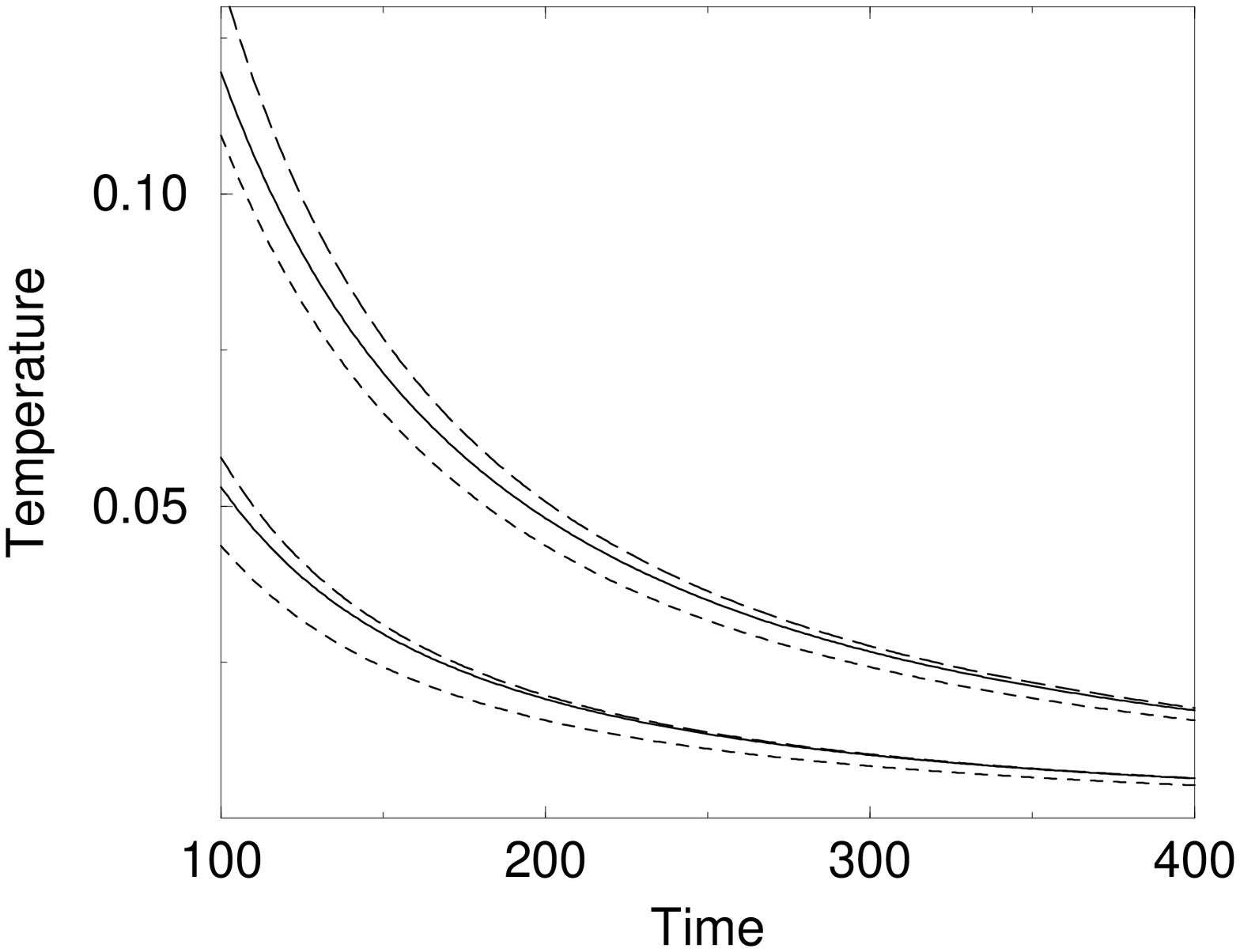,width=8cm}}
\centerline{\psfig{figure=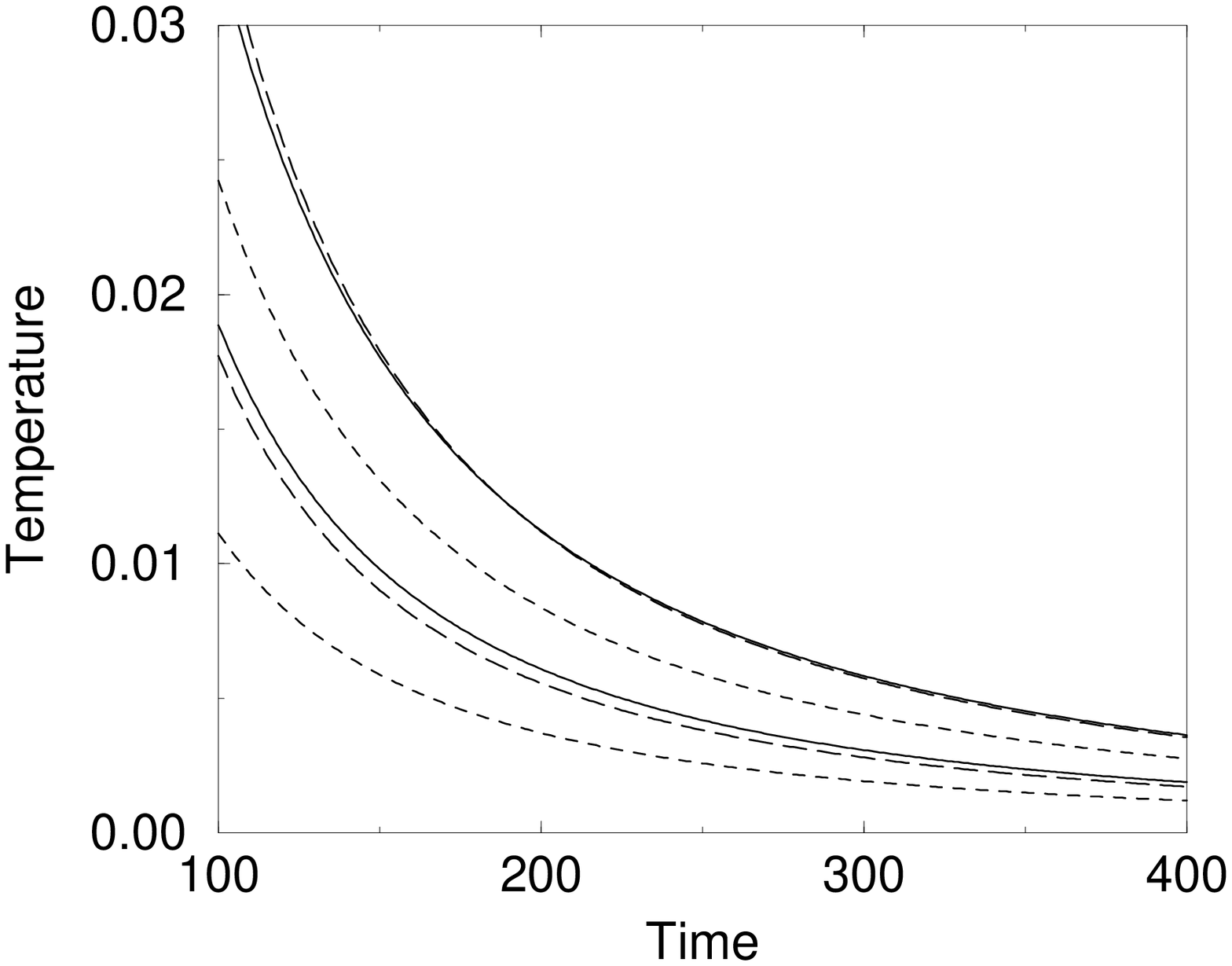,width=8cm}}
\centerline{\psfig{figure=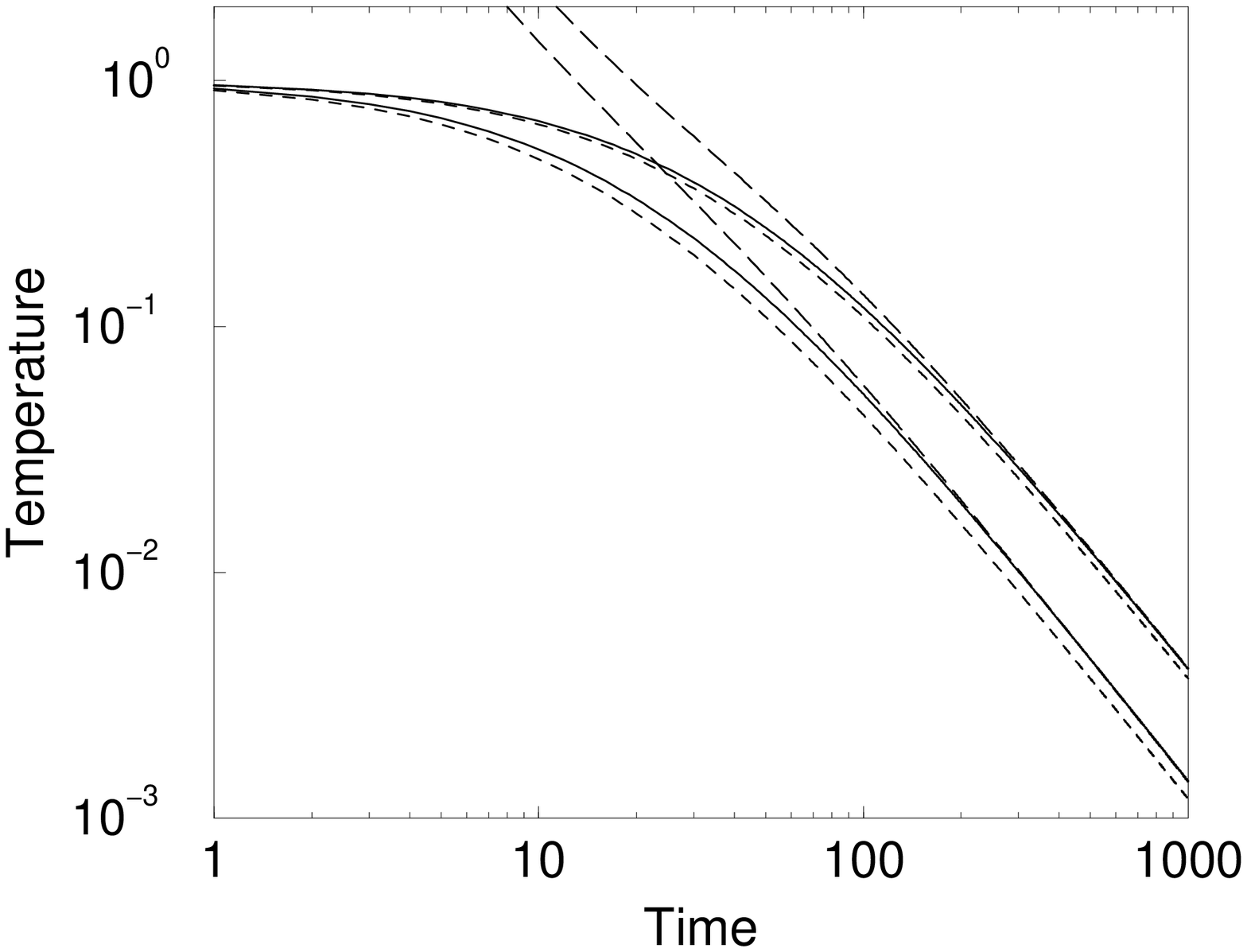,width=8cm}}
  \caption{Time-evolution of the reduced temperature, $u(t)=T(t)/T_0$.
The time is given in units of mean collisional time
$\tau_c(0)$. Solid line: numerical solution, short-dashed:
$u_0(t)=(1+t/\tau_0)^{-5/3}$ (zero-order theory),
long-dashed: $u(t)=u_0(t)+\delta \, u_1(t)$ (first-order theory).
(Top): for $\delta=0.05, 0.1$ (top to bottom); (Middle): $\delta=0.15, 0.25$ (top to bottom);
(Bottom): the same as (Top) but log-scale and larger ranges.}
  \label{fig:T}
\end{figure}
\end{minipage}

According to our analysis of the diffusion in granular gas of viscoelastic particles 
\cite{BrilliantovPoeschel:1998d}, the clustering is expected to be retarded, compared to 
the case of a constant $\epsilon$. Therefore, we may assume that for the time shown on the 
figures the granular gas is still in the regime of homogeneous cooling.

For larger values of $\delta$ the linear theory breaks down. Unfortunately, the
equations obtained for the second order approximation ${\cal O}(\delta^2)$ are
too complicated to be treated analytically. Hence, we studied them 
only numerically
(see Fig.~\ref{fig:a2time}). As compared to the case of small $\delta$, an additional intermediate regime
 in
the time-evolution of the velocity distribution is observed. The first ``fast'' stage
of evolution takes place, as before, on the time scale of few collisions per particle,
where maximal deviation from the Maxwellian distribution is achieved (Fig.~\ref{fig:a2time}). For
$\delta \geq 0.15$ these maximal values of $a_2$ are  positive. Then, on 
the second
stage (intermediate regime), which continues $10-100$ collisions, $a_2$ 
changes its
sign and reaches a maximal negative deviation. Finally, on the third, slow stage,
$a_2(t)$ relaxes to zero on the slow time-scale $\sim \tau_0$, just as for
small $\delta$. In Fig.~\ref{fig:a2time} we show the first stage of the
time evolution of $a_2(t)$ for systems  with large $\delta$. At a certain value of the dissipative parameter $\delta$ the behavior changes qualitatively, i.e. the system then reveals another time scale as discussed above.
\begin{minipage}{8cm}
\begin{figure}[htbp]
\centerline{\psfig{figure=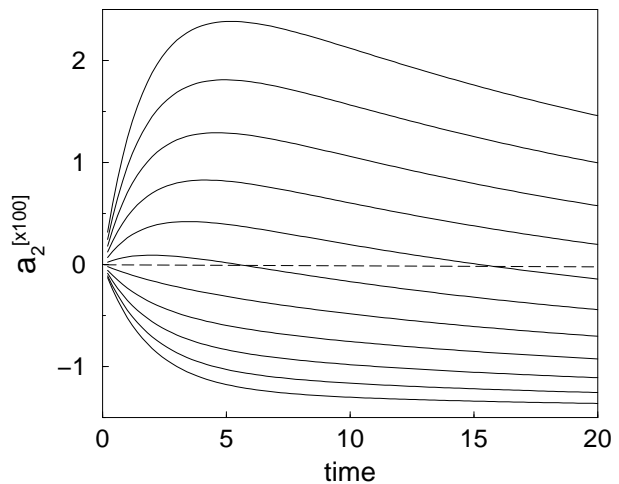,width=8cm}}
  \caption{Time dependence of the second coefficient of the Sonine polynomial expansion $a_2(t) \times 100 $.
Time is given in units of mean collisional time $\tau_c(0)$. $\delta=0.1, 0.11, 0.12, \ldots, 0.20$ (bottom to top).}
  \label{fig:a2time}
\end{figure}
\end{minipage}

Figure~\ref{fig:a2evo} shows the numerical solution of Eqs.~(\ref{genseteq1},\ref{genseteq2}) for the second Sonine coefficient $a_2(t)$ as a function of time. One can clearly distinguish the different stages of evolution of the velocity distribution function. 

Thus we conclude that for the case of not very small dissipative parameter $\delta$ the time
evolution  of the velocity distribution function (described on the level of the second 
coefficient of the Sonine polynomials expansion) exhibits very complicated nonmonotonic
behavior with few different regimes. Physically such behavior is caused by existing of 
an additional intrinsic time-scale, which describes the viscoelastic collision and by coupling 
of the evolution of the velocity distribution with the time-evolution of temperature.   
\begin{minipage}{8cm}
\begin{figure}[htbp]
\centerline{\psfig{figure=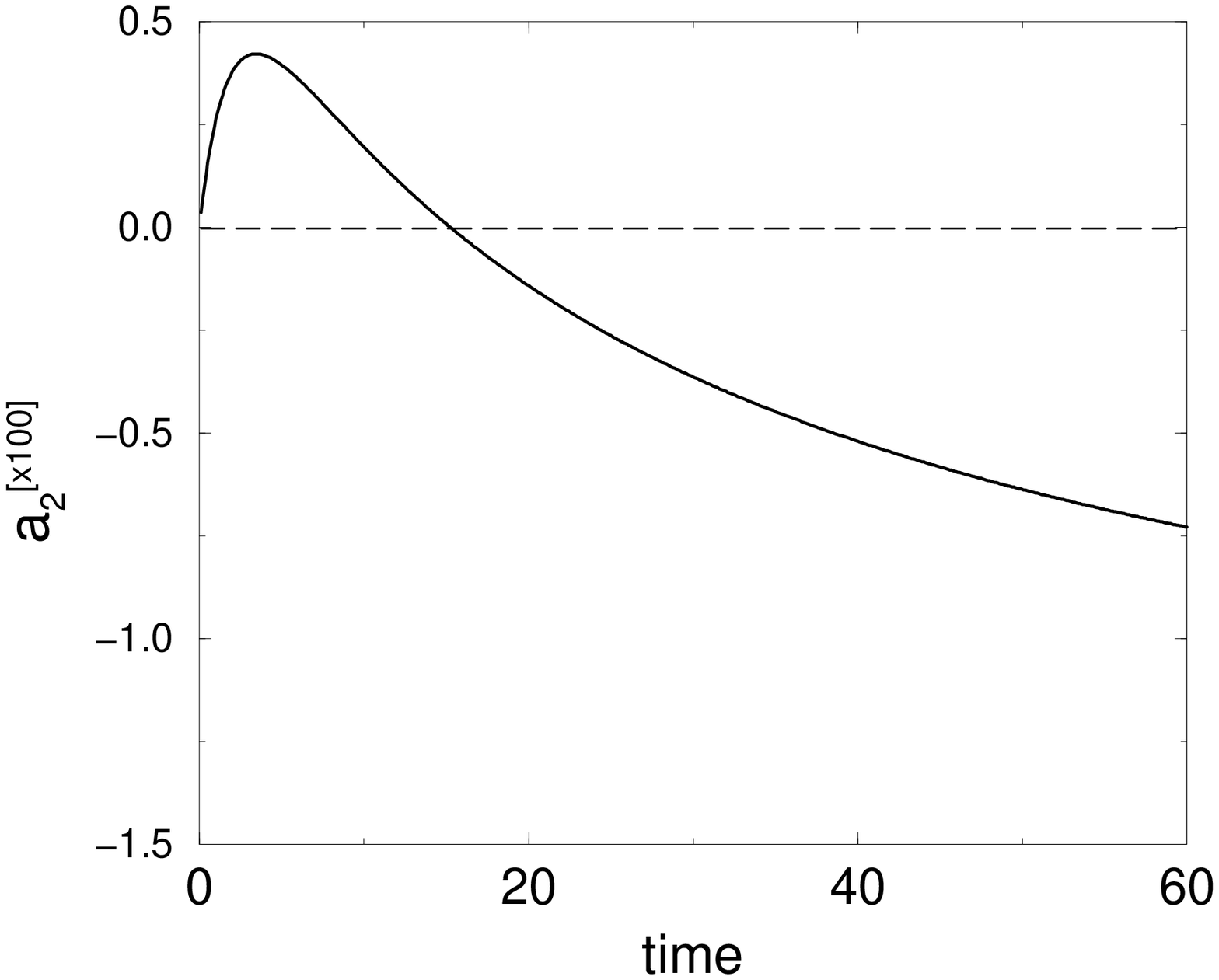,width=6cm}}
\centerline{\psfig{figure=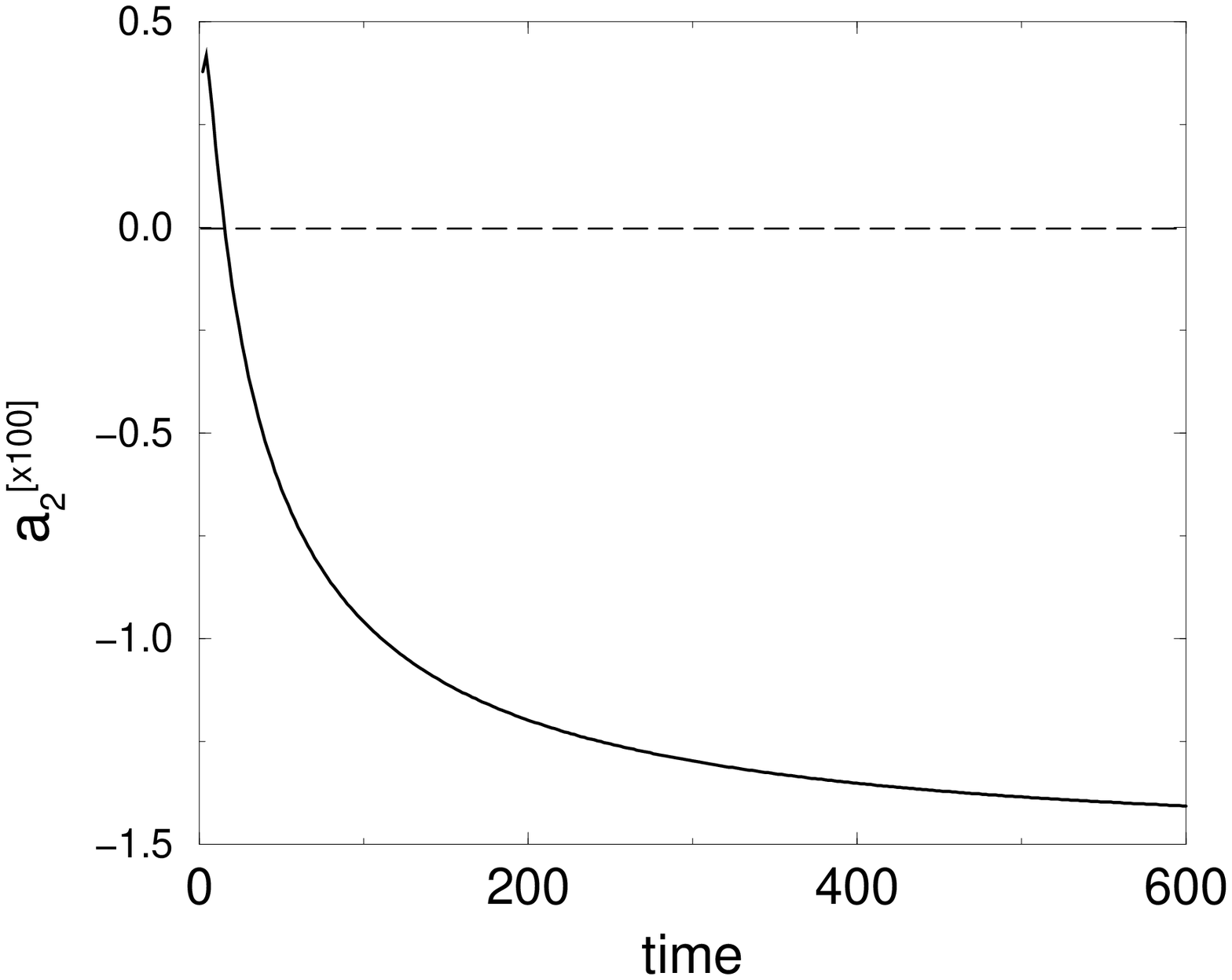,width=6cm}}
\centerline{\psfig{figure=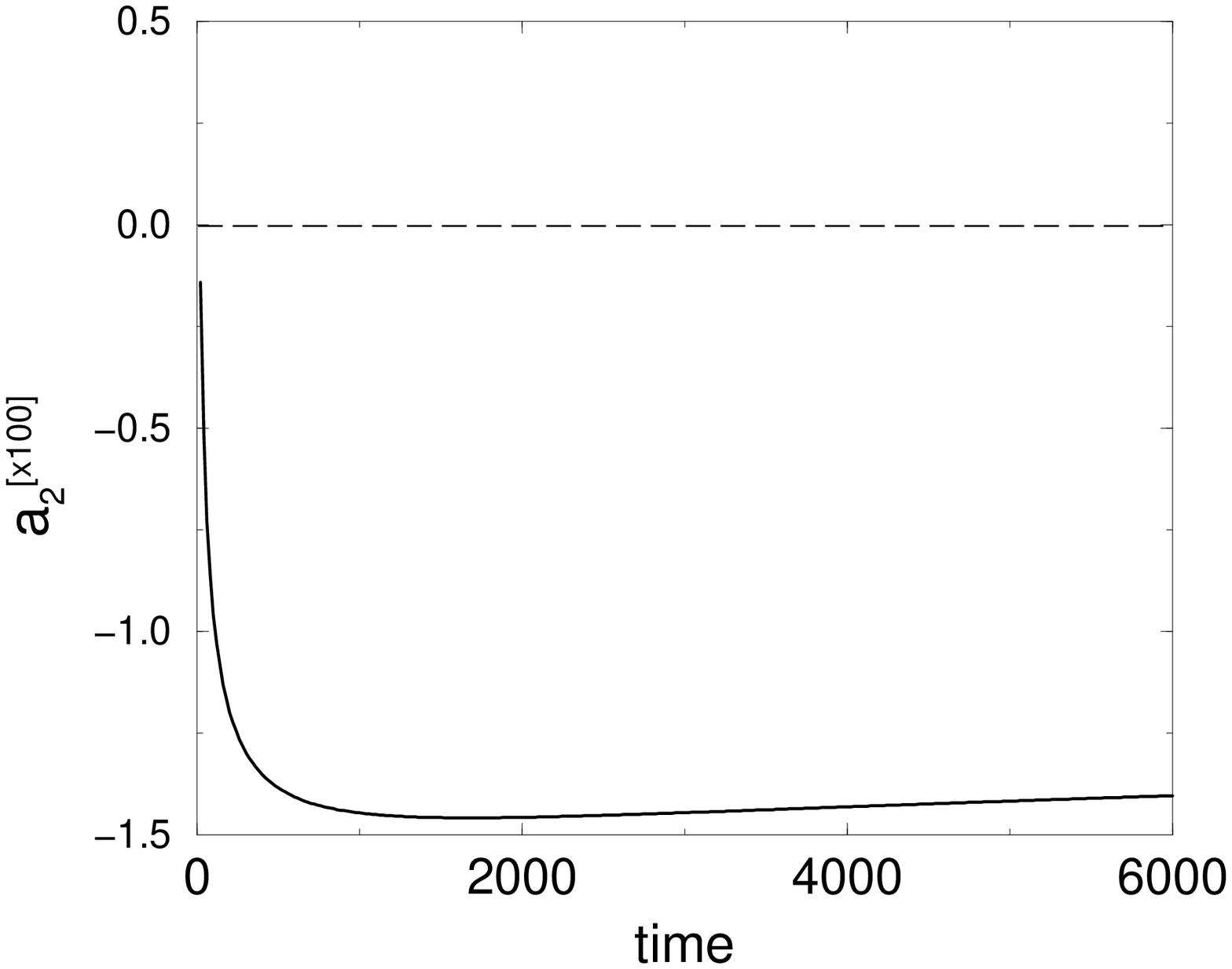,width=6cm}}
\centerline{\psfig{figure=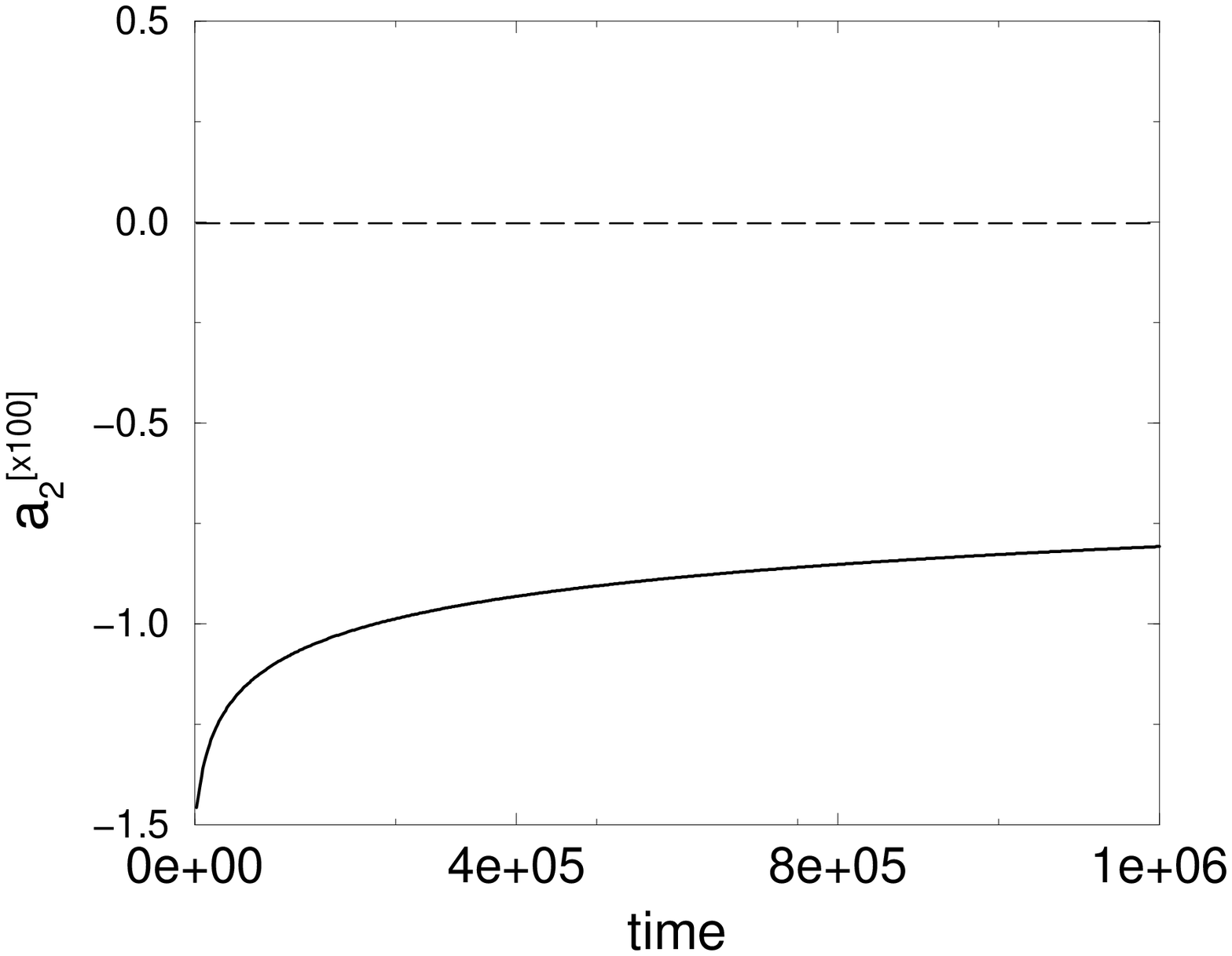,width=6cm}}
  \caption{The second Sonine coefficient $a_2$  for $\delta=0.16$ over time. The numerical solutions of Eqs.~(\ref{genseteq1},\ref{genseteq2}) show all stages of evolution discussed in the text.}
  \label{fig:a2evo}
\end{figure}
\end{minipage}
The analysis which has been performed up to now has been  addressed to the main part of the velocity 
distribution function. The most important component  of the distribution is still the 
Maxwellian, while deviations from this have been  quantified 
in terms of the Sonine polynomial expansion. For very large velocities however this is not
true and the Maxwellian distribution may not be used as a zero-order approximation. In 
the next section we address the problem of properties of the velocity distribution 
function for $v \gg v_0$.  

\section{High-velocity tail of the velocity-distribution function} 
\label{sec:Tail}

The high-velocity tail of the velocity distribution function in force-free granular 
gases was analyzed for the case of a constant restitution coefficient in 
Refs.~\cite{EsipovPoeschel:97,NoijeErnst:97}. It was shown in these studies that 
for large velocities, $c \gg 1$, the velocity distribution function behaves 
as $\tilde{f}(c) \sim \exp(- {\rm const} \cdot c)$, i.e. that the tail $c \gg 1 $ is 
overpopulated, as compared to the Maxwellian distribution $ \sim \exp(-c^2)$. 

Here we use the scheme of analysis proposed in Ref.~\cite{EsipovPoeschel:97}.
The same arguments as in \cite{EsipovPoeschel:97,NoijeErnst:97}, lead to 
a conclusion  that the gain term of the collisional integral $\tilde{I}$ may be 
neglected for $c \gg 1$ with respect to the loss term, which does not 
depend on the restitution coefficient. Thus, following 
\cite{EsipovPoeschel:97,NoijeErnst:97}, we approximate the collision integral 
as 
\begin{equation}
\tilde{I}\left( \tilde{f}, \tilde{f} \right) \approx - \pi c \tilde{f}(\vec{c}, t)
\end{equation}
and write for $c \gg 1$ the kinetic equation (\ref{geneqveldis})  as 
\begin{equation}
\label{geneqveldis1}
\frac{\mu_2}{3} 
c \frac{\partial}{\partial c}  \tilde{f}(\vec{c}, t) +
B^{-1} \frac{\partial}{\partial t} \tilde{f}(\vec{c}, t) 
\approx - \pi c \tilde{f}(\vec{c}, t)
\end{equation}

If one would use the expansion (\ref{genSoninexp}) 
(with coefficients $a_p$ for $p>2$ discarded)
to substitute it into 
Eq. (\ref{geneqveldis1}) one would obtain for the {\em second}  term in the 
left-hand side of (\ref{geneqveldis1}) at $c \gg 1$:
\begin{equation}
\label{Bdfdt}
B^{-1} \frac{\partial \tilde{f}}{\partial t} =
\frac43 \left[ \mu_2(1+a_2)-\frac15 \mu_4 \right] \phi(c) S_2(c^2) 
\sim c^4 e^{-c^2}
\end{equation}
where we have used the relation 
\begin{equation}
\partial \tilde{f} /\partial t = \dot{a}_2 \phi(c)S_2(c^2) \, , 
\end{equation} 
$\dot{a}_2$ according to Eq. (\ref{eqa2}), and the definition 
(\ref{Soninfewfirst}) of $S_2(c^2)$, 
which shows that $S_2(c^2) \sim c^4$ at $c \gg 1$.
We also take into account that $\mu_2$, $\mu_4$ and $a_2$ do not depend on $c$. 
For the first term in the left-hand side of (\ref{geneqveldis1}) and for the   
right-hand side of (\ref{geneqveldis1}) this substitute yields correspondingly 
in the same limit $c \gg 1$:
\begin{eqnarray}
\label{cdfdc}
c \, \frac{\partial \tilde{f} }{\partial c}  &\sim& -c^6 e^{-c^2} \\
\label{cf}
c \,\tilde{f}   &\sim& c^5 e^{-c^2}\,.
\end{eqnarray} 
From the last Eqs. (\ref{Bdfdt}), (\ref{cdfdc}) and (\ref{cf}) follows that 
although all terms in the Eq. (\ref{geneqveldis1}) have the same factor 
$e^{-c^2}$, the exponents of the power of $c$ of the prefactor are different 
for all terms. This means inconsistency of the substitute 
(\ref{genSoninexp}), with $a_p$ for $p>2$ discarded, 
for $c \gg 1$. Similarly, it may be shown that such 
inconsistency appears for  any order of the Sonine polynomial expansion. 
Indeed, using the Sonine polynomial expansion  (\ref{genSoninexp}) up to 
(arbitrary) order $n$, yields the estimate $\sim c^{(2n+2)}e^{-c^2}$ for the 
first term and $\sim c^{2n}e^{-c^2}$ for the second term in the 
left-hand side of (\ref{geneqveldis1}), while for the 
right-hand side of (\ref{geneqveldis1}) one obtains $\sim c^{(2n+1)}e^{-c^2}$.

The exponential ansatz 
\begin{equation}
\label{expansatz}
\tilde{f}(\vec{c}, t) \sim \exp \left\{ -\varphi(t) \cdot c \right\}
\end{equation}
for the kinetic equation (\ref{geneqveldis1}) occurs, however, to be 
self-consistent for $c \gg 1$. Substituting this into Eq. (\ref{geneqveldis1}) one 
finds that the function $\varphi(t)$ in (\ref{expansatz}) must satisfy 
\begin{equation}
\label{dvarphidt}
\dot{\varphi}+\frac13 \mu_2 B \, \varphi = \pi B \,,
\end{equation}
where the time-dependence of $B$ is given by Eq. (\ref{BT}) and $\mu_2$ depends on 
time via $a_2(t)$ according to (\ref{MU2A}). 
In linear with respect to $\delta$ approximation $a_2 \sim \delta$,  and therefore,  
according to (\ref{MU2A}), (\ref{A1A6}), $\mu_2(t) \simeq \delta^{\prime} \omega_0$.
Using then the definition (\ref{deltaprime}) of $\delta^{\prime}$ and expression 
(\ref{BT}) for $B(t)$, one obtains
\begin{eqnarray}
\mu_2(t) B(t)&=&\frac52\, \tau_0^{-1}u^{3/5}(t)\nonumber\\
\qquad B(t)&=&\frac{\tau_c(0)^{-1}}{\sqrt{8 \pi}}u^{1/2}(t)
\label{MU2B}
\end{eqnarray}
with $\tau_0$ being defined by Eq. (\ref{tau0}). With Eq. (\ref{T(t)del1}) for the time-dependence of temperature in this approximation, Eq. (\ref{dvarphidt}) reads
\begin{equation}
\label{dvarphidt1}
\dot{\varphi}+\frac{5}{6\tau_0}\left(1+\frac{t}{\tau_0} \right)^{-1} \varphi
=\sqrt{\frac{\pi}{8}} \tau_c(0)^{-1} \left(1+\frac{t}{\tau_0} \right)^{-5/6}\,.
\end{equation}
Substituting the ansatz $\varphi \sim (1+t/\tau_0)^{\nu}$ we find the exponent, 
$\nu =1/6$ and the prefactor, so we arrive at the final result
\begin{equation}
\label{resvarphi}
\varphi(t)=b \delta^{-1}\left(1+\frac{t}{\tau_0} \right)^{1/6}
\end{equation}
with
\begin{eqnarray}
b=\sqrt{\frac{\pi}{2}}\left(\frac{5}{16q_0}\right)=\frac{5^{7/5}}{2^{3/2}\Gamma(3/5)}
=2.25978\ldots
\end{eqnarray}
Thus, the velocity distribution function reads for $c \gg 1$:
\begin{equation}
\label{veldiscgg1}
\tilde{f}(\vec{c}, t) \sim 
\exp 
\left[ - \frac{b}{ \delta}\,c\left(1+\frac{t}{\tau_0} \right)^{1/6} \right]\, .
\end{equation}
Note that the obtained expression (\ref{veldiscgg1}) refers only for times $t \gg \tau_c(0)$, 
when the deviations from the Maxwellian distribution are already well developed; 
it is not applicable for the transient times $t \sim  \tau_c(0)$.

As one can see from Eq. (\ref{veldiscgg1}) the overpopulation 
(with respect to Maxwellian distribution) 
of the high-velocity tail
decreases with time on the same time-scale $\sim \tau_0$ as $a_2(t)$, i.e., the 
velocity distribution in the system approaches the Maxwellian. 
However, it should be noted that the above considerations are valid as 
long as the overpopulation in the tail is 
significant to make the gain term in collision integral be negligible as 
compared to the loss term. 

\section{conclusion}
We studied the velocity distribution in a homogeneously cooling granular gas of viscoelastic particles which implies an impact velocity dependent restitution coefficient. We observed that contrary to the case of the 
constant restitution coefficient, the distribution function may not be represented 
in a simple scaling form, where the time dependence of the function occurs only 
via the time dependence of the temperature. The dependence of the 
restitution coefficient on the impact-velocity  causes a time dependence of the coefficients of the Sonine polynomials expansion, which describes the deviation of the velocity distribution 
from the Maxwellian. 

We analyzed the time evolution of the temperature and of the second coefficient of 
the Sonine polynomials expansion $a_2$. Contrary to the case of the constant restitution 
coefficient, the evolution of temperature is coupled now with the time 
evolution of $a_2$. 

For small values of the dissipative parameter 
$\delta$ we developed an analytical theory for the time evolution of the temperature 
of the granular gas and for  the coefficient of the Sonine polynomials expansion $a_2$;
the case of larger $\delta$ was studied numerically. 
We observed a complicated nonmonotonic
time-behavior of the  coefficient $a_2$. For small values of the dissipative parameter 
$\delta$ we detected two different stages in its time
evolution: a first fast stage, which develops on the time scale of the mean-collision
time $\tau_c$, and the second, slow stage on the time scale $ \sim \tau_0 \gg \tau_c$, 
on which the temperature of the granular gas changes. In  the fast stage a maximal deviation 
from the Maxwellian distribution is achieved and then the deviation relaxes to zero 
during the second slow stage. Our numerical results agree well with the predictions
of the analytical theory for small $\delta$.

When $\delta$ is not small, a much more complicated time behavior 
of the coefficient $a_2$ has been revealed. In addition to the two stages of evolution which 
have been observed for the case of small dissipative parameter, a regime of intermediate 
relaxation has been detected. Physically such complicated behavior is caused by 
an additional intrinsic time-scale, which describes the viscoelastic collision and by coupling 
of the evolution of the velocity distribution with the time-evolution of temperature.   

We also analyzed the high-velocity tail of the velocity distribution for the case of the 
impact-velocity dependent restitution coefficient for viscoelastic particles. 
We found the same exponential 
overpopulation of the tail as for the constant restitution coefficient. However, contrary 
to the latter case where the overpopulation of the tail persists with time, it
decreases for the impact dependent restitution coefficient and the velocity distribution tends to the Maxwellian as the system evolves. 

\bigskip
The authors want to thank I. Goldhirsch and M. H. Ernst for helpful discussions. 
The work was supported by Deutsche Forschungsgemeinschaft 
through grant Po 472/3-2.

\appendix

\section{Derivation of Eq. (\ref{CHI})}

The change of the particle velocities due to the inverse collision is described by
\begin{eqnarray}
\label{invcoll}
&&\vec{v}_1=\vec{v}_1^{**}-
\frac12 \left[1+\epsilon\left(g^{**}\right)\right] g^{**} \vec{e} \nonumber \\
&&\vec{v}_2=\vec{v}_2{**}+
\frac12 \left[1+\epsilon\left(g^{**}\right)\right] g^{**} \vec{e} 
\end{eqnarray}
where we introduce the normal relative velocity $g^{**} \equiv \vec{v}_{12}^{**} \cdot \vec{e}$ and where 
\begin{equation}
\label{epsinv}
\epsilon\left(g^{**}\right)=1-C_1 A \alpha^{2/5} \left|g^{**}\right|^{1/5}
+C_2A^2 \alpha^{4/5}\left|g^{**}\right|^{2/5}  \mp \cdots
\end{equation}
according to the viscoelastic character of the particles (see Eq.~(\ref{epsC1C2})). Equations (\ref{invcoll}) and (\ref{epsinv}) imply the conservation of momentum
\begin{equation}
\label{impcons}
\vec{v}_{1}+\vec{v}_{2}=\vec{v}_{1}^{**}+\vec{v}_{2}^{**}
\end{equation}
and the relation:
\begin{equation}
\label{g**g}
g=-\epsilon\left(g^{**}\right)g^{**}
\end{equation}
with $g \equiv \vec{v}_{12} \cdot \vec{e}$. Using  
 $C_2=\frac35C_1^2$ (Eq. (\ref{C1C22})), one can also write:
\begin{equation}
\label{g**g1}
g^{**}=g\left[ 1+C_1 A \alpha^{2/5} \left|g\right|^{1/5} + 
\frac35 \left(C_1 A \alpha^{2/5}\right)^2\left|g\right|^{2/5} \mp \cdots \right]
\end{equation}
We use Eq. (\ref{g**g1}) to find the relation 
between the length of the collisional cylinders, $\left|g\right|dt$ and $\left|g^{**}\right|dt$, when the 
transformation of variables $\vec{v}_{1}^{\,**}, \vec{v}_{2}^{\, **} \to \vec{v}_{1}\vec{v}_{2}$
is made in the collisional integral. One also needs the Jacobian of this transformation. 
To calculate this, it is convenient to chose the coordinate axis $Z$ along the inter-center 
vector $\vec{e}$, i.e.,
\begin{equation}
g \equiv v_{12, z} \equiv v_{1,z}-v_{2,z} \,.
\end{equation}
Then from Eqs. (\ref{invcoll}) follows:
\begin{eqnarray}
\label{vxvyvz}
&&v_{1,x}^{**}=v_{1,x};  \qquad v_{1,y}^{**}=v_{1,y}; \\
&&v_{2,x}^{**}=v_{2,x};  \qquad v_{2,y}^{**}=v_{2,y};  \nonumber \\
&&v_{1,z}^{**}=v_{1,z}+\frac12\left(g^{**}+g\right); \nonumber \\
&&v_{2,z}^{**}=v_{2,z}-\frac12\left(g^{**}+g\right)\,, \nonumber 
\end{eqnarray}
where the value of $g^{**}$ is expressed in terms of $g$ (i.e. in terms of $v_{1,z}$ and $v_{2,z}$)
according to Eq. (\ref{g**g1}). Thus, Eqs. (\ref{vxvyvz}) explicitly express all  
components of the inverse-collision velocities $\vec{v}_{1}^{**}$,  $\vec{v}_{2}^{**}$ 
in terms of $\vec{v}_{1}$,  $\vec{v}_{2}$. Straightforward calculations yield for the 
Jacobian
\begin{eqnarray}
d\vec{v}_{1}^{\, **} d\vec{v}_{2}^{\, **}&=&
\left[1+\frac65 C_1 A \alpha^{2/5} \left|g\right|^{1/5} + \right.\nonumber\\
&+&\left.\frac{21}{25} \left(C_1 A \alpha^{2/5}\right)^2\left|g\right|^{2/5} \mp \cdots \right]\,.
\label{Jacobian}
\end{eqnarray}
Combining (\ref{Jacobian})  with Eq. (\ref{g**g1}) which relates the lengths of 
collisional cylinders, one arrives at the factor $\chi$, Eq. (\ref{CHI}), in the 
collisional integral. 

\end{multicols}

\section{Derivation of the moments $\mu_p$ (Eqs. (\ref{MU2A},\ref{MU4A}))}
To calculate the moments 
\begin{equation}
\label{mupa21}
\mu_p=-\frac12 \int d\vec{c}_1\int d\vec{c}_2 \int d\vec{e} 
\Theta\left(-\vec{c}_{12} \cdot \vec{e}\right) \left|\vec{c}_{12} \cdot \vec{e}\right| \phi\left(c_1\right) \phi\left(c_2\right) 
\left\{1+a_2\left[S_2\left(c_1^2\right)+S_2\left(c_2^2\right) \right] + a_2^2\,S_2\left(c_1^2\right)S_2\left(c_2^2\right) \right\}
\Delta \left(c_1^p+c_2^p\right) 
\end{equation}
it is convenient to use the center of mass velocity $\vec{C}$ and relative velocity 
$\vec{c}_{12}$ such that 
\begin{equation}
\label{Cc12viac1c2}
\vec{c}_{1}=\vec{C}+\frac12 \vec{c}_{12}~, \qquad \vec{c}_{2}=\vec{C}-\frac12 \vec{c}_{12}
\end{equation}
The Jacobian of the transformation (\ref{Cc12viac1c2}) is equal to unity and the product 
$\phi\left(\vec{c}_{1}\right) \phi\left(\vec{c}_{2}\right)$ transforms into
\begin{equation}
\label{MaxtoMax}
\phi\left(\vec{c}_{1}\right) \phi\left(\vec{c}_{2}\right) \to 
\frac{1}{(2 \pi)^{3/2}} \exp \left(-\frac12 c_{12}^2 \right)
\left(\frac{2}{\pi} \right)^{3/2} \exp \left( -2C^2 \right) \equiv 
\phi\left(\vec{c}_{12}\right) \phi\left(\vec{C}\right)\,.
\end{equation}
In terms of the variables $\vec{C}$ and $\vec{c}_{12}$ the quantity 
$\left[ S_2\left(c_1^2\right)+S_2\left(c_2^2\right) \right]$ in Eq. (\ref{mupa21}) may be written as
\begin{equation}
\label{S2S2Cc12}
\left[ S_2\left(c_1^2\right)+S_2\left(c_2^2\right) \right]=
C^4+\left(\vec{C} \cdot \vec{c}_{12} \right)^2 +\frac{1}{16}c_{12}^4
+\frac12  C^2 c_{12}^2-5 C^2-\frac54 c_{12}^2 +\frac{15}{4}\, .
\end{equation}
For $S_2\left(c_1^2\right)S_2\left(c_2^2\right)$ we obtain
\begin{equation}
\label{S2xS2Cc12}
S_2\left(c_1^2\right)S_2\left(c_2^2\right)=K_1+K_2+K_3+K_4
\end{equation}
where
\begin{eqnarray}
\label{K1}
K_1&=&\frac14 C^8 -\frac52 C^6 +\frac{65}{8} C^4-\frac{75}{8} C^2 \\
K_2&=&\frac{1}{1024} c_{12}^8-\frac{5}{128} c_{12}^6 +\frac{65}{128} c_{12}^4
-\frac{75}{32} c_{12}^2 \\
K_3&=&\frac{3}{32}C^4c_{12}^4 +\frac14 C^6c_{12}^2 +\frac{1}{64}C^2c_{12}^6 -\frac{15}{8}C^4c_{12}^2 
-\frac{15}{32}C^2c_{12}^4 +\frac{65}{16}C^2c_{12}^2 \\
K_4&=&\frac14 \left(\vec{C} \cdot \vec{c}_{12} \right)^4 
-\frac14 C^2 \left(\vec{C} \cdot \vec{c}_{12} \right)^2 c_{12}^2 
-\frac12 C^4 \left(\vec{C} \cdot \vec{c}_{12} \right)^2
-\frac{1}{32} \left(\vec{C} \cdot \vec{c}_{12} \right)^2 c_{12}^4 
+\frac52 C^2 \left(\vec{C} \cdot \vec{c}_{12} \right)^2 \nonumber\\
&+&\frac58 \left(\vec{C} \cdot \vec{c}_{12} \right)^2 c_{12}^2
-\frac{35}{8} \left(\vec{C} \cdot \vec{c}_{12} \right)^2 +\left(\frac{15}{8} \right)^2
\end{eqnarray}
For the quantities $\Delta\left(c_1^p+c_2^p\right)$ ($p=2,4$) we find
\begin{equation}
\label{Deltac1c2p2}
\Delta\left(c_1^2+c_2^2\right)=-\frac12\left(\vec{c}_{12} \cdot \vec{e} \right)^2 \left(1-\epsilon^2\right)
\end{equation}
and
\begin{eqnarray}
\label{Deltac1c24}
\Delta\left(c_1^4+c_2^4\right)&=&2\left(1+\epsilon\right)^2\left(\vec{c}_{12} \cdot \vec{e} \right)^2
\left(\vec{C} \cdot \vec{e} \right)^2 +
\frac18 \left(1-\epsilon^2\right)^2 \left(\vec{c}_{12} \cdot \vec{e} \right)^4-\frac14 \left(1-\epsilon^2\right) \left(\vec{c}_{12} \cdot \vec{e} \right)^2 \vec{c}_{12}^2 -
\vec{C}^2 \left(1-\epsilon^2\right) \left(\vec{c}_{12} \cdot \vec{e} \right)^2 -\nonumber \\
&-& 4 \left(1+\epsilon\right) \left(\vec{C} \cdot \vec{c}_{12} \right) \left(\vec{C} \cdot \vec{e} \right) 
\left(\vec{c}_{12} \cdot \vec{e} \right)\,.
\end{eqnarray}
Substituting (\ref{S2S2Cc12}), (\ref{S2xS2Cc12}), (\ref{Deltac1c2p2}) and 
(\ref{Deltac1c24}) into (\ref{mupa21}) and using the expansions

\begin{eqnarray}
\label{epsexpan}
\left(1-\epsilon^2\right)&=&2C_1\delta^{\prime}(t) \left|\vec{c}_{12} \cdot \vec{e} \right|^{1/5}-
\frac{11}{5}C_1^2 \delta^{\prime \, 2}(t) \left|\vec{c}_{12} \cdot \vec{e} \right|^{2/5} + \cdots \\
\left(1+\epsilon\right)^2&=&4\left[1-C_1 \delta^{\prime}(t) \left|\vec{c}_{12} \cdot \vec{e} \right|^{1/5}
+\frac{17}{20} C_1^2 \delta^{\prime \, 2}(t) \left|\vec{c}_{12} \cdot \vec{e} \right|^{2/5} + \cdots \right] \\
\left(1-\epsilon^2\right)^2&=&4C_1^2 \delta^{\prime \, 2}(t) \left|\vec{c}_{12} \cdot \vec{e} \right|^{2/5} + \cdots 
\end{eqnarray}
one observes that $\mu_2$ and $\mu_4$ may be expressed in terms of the 
basic integrals
\begin{equation}
\label{basicint}
J_{k,l,m,n,p,\alpha}=
\int d \vec{c}_{12} \int d \vec{C} \int d\vec{e} 
\Theta\left(-\vec{c}_{12} \cdot \vec{e}\right) \left|\vec{c}_{12} \cdot \vec{e}\right|^{1+\alpha} 
\phi\left(c_{12}\right) \phi(C)
C^k c_{12}^l \left(\vec{C} \cdot \vec{c}_{12} \right)^m \left(\vec{C} \cdot \vec{e} \right)^n 
\left(\vec{c}_{12} \cdot \vec{e} \right)^p 
\end{equation}
Namely, one has for $\mu_2$:
\begin{equation}
\label{mu2viaJ}
\mu_2=\frac12 \, \delta^{\prime} \, C_1 
\left[J_{0,0,0,0,2,1/5}+a_2 L\left(\frac15\right)+
a_2^2M\left(\frac15\right) \right] 
-\frac{11}{20} \, \delta^{\prime \,2} \, C_1^2
\left[J_{0,0,0,0,2,2/5}+a_2 L\left(\frac25\right)+
a_2^2M\left(\frac25\right) \right] 
\end{equation}
where we define
\begin{equation}
L(\alpha)=J_{4,0,0,0,2,\alpha}+J_{0,0,2,0,2,\alpha}+\frac{1}{16}J_{0,4,0,0,2,\alpha} 
+\frac12 J_{2,2,0,0,2,\alpha}-5J_{2,0,0,0,2,\alpha} -\frac54J_{0,2,0,0,2,\alpha}
+\frac{15}{4}J_{0,0,0,0,2,\alpha}
\label{defL}
\end{equation}
and 
\begin{eqnarray}
\label{defM}
M(\alpha)&=&
\frac14 J_{8,0,0,0,2,\alpha} 
-\frac52 J_{6,0,0,0,2,\alpha}
+\frac{65}{8} J_{4,0,0,0,2,\alpha} 
-\frac{75}{8} J_{2,0,0,0,2,\alpha}
+\frac{1}{1024} J_{0,8,0,0,2,\alpha}
-\frac{5}{128} J_{0,6,0,0,2,\alpha}\nonumber \\
&+&\frac{65}{128} J_{0,4,0,0,2,\alpha}
-\frac{75}{32} J_{0,2,0,0,2,\alpha} 
+\frac{3}{32} J_{4,4,0,0,2,\alpha}
+\frac14 J_{6,2,0,0,2,\alpha}
+\frac{1}{64} J_{2,6,0,0,2,\alpha} 
-\frac{15}{8} J_{4,2,0,0,2,\alpha} \nonumber  \\
&+&\frac{65}{16} J_{2,2,0,0,2,\alpha}
+\frac14 J_{0,0,4,0,2,\alpha}
-\frac14 J_{2,2,2,0,2,\alpha}
-\frac12 J_{4,0,2,0,2,\alpha}
-\frac{1}{32} J_{0,4,2,0,2,\alpha}
+\frac52 J_{2,0,2,0,2,\alpha} \nonumber  \\
&+&\frac58 J_{0,2,2,0,2,\alpha}
-\frac{35}{8} J_{0,0,2,0,2,\alpha}
+\left(\frac{15}{8} \right)^2 J_{0,0,0,0,2,\alpha}
\end{eqnarray}
The basic integrals may be calculated (details are given in 
Appendix C) and the following 
expressions are obtained:

\begin{equation}
\label{Jn0}
J_{k,l,m,0,p,\alpha}=
\frac{(-1)^p \cdot 8 \cdot 2^{\frac{-k+l+p+\alpha-1}{2}}}{(p+\alpha+2)(m+1)}
\left[1-(-1)^{m+1} \right]
\Gamma\left( \frac{k+m+3}{2} \right)
\Gamma\left( \frac{l+m+p+\alpha+4}{2} \right)
\end{equation}
for $n=0$,
\begin{equation}
\label{Jn1}
J_{k,l,m,1,p,\alpha}=
\frac{(-1)^{p+1} \cdot 4 \cdot 2^{\frac{-k+l+p+\alpha}{2}}}{(p+\alpha+3)(m+2)}
\left[1-(-1)^{m} \right]
\Gamma\left( \frac{k+m+4}{2} \right)
\Gamma\left( \frac{l+m+p+\alpha+4}{2} \right)
\end{equation}
for $n=1$ and 

\begin{equation}
\label{Jn2}
J_{k,l,m,2,p,\alpha}=
\frac{(-1)^p \cdot 4 \cdot 2^{\frac{-k+l+p+\alpha-1}{2}}}{(p+\alpha+2)(m+1)}
\left[1-(-1)^{m+1} \right]
\Gamma\left( \frac{k+m+5}{2} \right)
\Gamma\left( \frac{l+m+p+\alpha+4}{2} \right)
\left[\frac{p+\alpha+1}{m+3}+\frac{1}{m+1} \right]
\end{equation}
for $n=2$.

When we compare Eqs. (\ref{mu2viaJ}) and (\ref{MU2A}) for $\mu_2$ and use relations
(\ref{Jn0}), (\ref{Jn1}) and (\ref{Jn2}) for the basic integrals, we find 
\begin{eqnarray}
{\cal A}_1&=&\frac12 C_1 J_{0,0,0,0,2,1/5} =2 \sqrt{2 \pi} 2^{1/10} \Gamma \left (\frac{21}{10} \right)C_1=6.48562 \ldots \equiv \omega_0\\
{\cal A}_4&=&\frac{11}{20}C_1^2 J_{0,0,0,0,2,2/5}=\sqrt{2 \pi} 2^{1/5} \Gamma \left (\frac{16}{5} \right)C_1^2=9.28569 \ldots \equiv \omega_1\,.
\end{eqnarray}
Computing from the basic integrals $L(\alpha)$ and $M(\alpha)$ for 
$\alpha=\frac15$ and $\alpha= \frac25$, and using the relation 
for the Gamma-function 
$\Gamma(x+1)=x\Gamma(x)$, we obtain 
\begin{eqnarray}
{\cal A}_2&=&\frac12 C_1L \left(\frac15\right)=\frac{6}{25}\omega_0\\
{\cal A}_5&=&\frac{11}{20}C_1^2L\left(\frac25\right)=\frac{119}{400}\omega_1\\
{\cal A}_3&=&\frac12C_1M\left(\frac15\right)=\frac{21}{2500} \omega_0\\
{\cal A}_6&=&\frac{11}{20}C_1^2M\left(\frac25\right)=\frac{4641}{640000}\omega_1.
\end{eqnarray}

Similar calculations may be performed for $\mu_4$  and yield Eq. (\ref{MU4A}) with the 
coefficients ${\cal B}_k$ expressed in terms of the basic integrals:
\begin{eqnarray}
{\cal B}_1&=&4\left( J_{0,0,1,1,1,0}-J_{0,0,0,2,2,0})=4(\sqrt{2\pi}-\sqrt{2\pi} \right)=0 \\
{\cal B}_2&=&-4\left(\frac{1}{16}J_{0,4,0,2,2,0}-J_{4,0,1,1,1,0}+J_{0,0,2,2,2,0} - J_{0,0,3,1,1,0}+J_{4,0,0,2,2,0}-\frac{1}{16}J_{0,4,1,1,1,0}+\frac12 J_{2,2,0,2,2,0}\right.\nonumber \\
&-&\frac12 J_{2,2,1,1,1,0} -\frac54 J_{0,2,0,2,2,0} +\frac54 J_{0,2,1,1,1,0}-5 J_{2,0,0,2,2,0}+5J_{2,0,1,1,1,0} \left. +\frac{15}{4}J_{0,0,0,2,2,0}-\frac{15}{4}J_{0,0,1,1,1,0} \right)\nonumber\\
&=&4\sqrt{2\pi} \\
{\cal B}_4&=&4C_1 \left( J_{0,0,0,2,2,1/5}-\frac12 J_{0,0,1,1,1,1/5} 
+\frac{1}{16}J_{0,2,0,0,2,1/5} +\frac14 J_{2,0,0,0,2,1/5}\right) =\frac{56}{5}\sqrt{2 \pi} 2^{1/10} 
\Gamma \left (\frac{21}{10} \right)C_1=
\frac{28}{5} \omega_0 \\
{\cal B}_7&=& C_1^2\left( \frac{11}{10} J_{2,0,0,0,2,2/5}+
\frac{11}{40} J_{0,2,0,0,2,2/5} +
\frac{17}{5} J_{0,0,0,2,2,2/5}+
 \frac14 J_{0,0,0,0,4,2/5}
-\frac65 J_{0,0,1,1,1,2/5} \right) \nonumber\\
&=&\frac{77}{10}\sqrt{2 \pi} 2^{1/5} \Gamma \left( \frac{16}{5} \right)C_1^2=\frac{77}{10}\omega_1
\end{eqnarray}
We do not give the expressions for the other few coefficients ${\cal B}_k$ in terms of the 
basic integrals, since they are too much cumbersome to be written explicitly. 
Computations of these is straightforward and yields the result:
\begin{eqnarray}
{\cal B}_3&=&\frac{1}{8}\sqrt{2 \pi}\\
{\cal B}_5&=&\frac{1806}{250}\omega_0\\
{\cal B}_6&=&\frac{567}{12500}\omega_0\\
{\cal B}_8&=&\frac{149054}{13750}\omega_1\\
{\cal B}_9&=&\frac{348424}{5500000}\omega_1
\end{eqnarray}

\begin{multicols}{2}
\section{Calculations of the basic integrals 
$J_{k,l,m,n,p,\alpha}$}

In this appendix we give some details for the calculations of the basic integrals 
$J_{k,l,m,n,p,\alpha}$. We need only integrals for $n=0, 1,2$. 
Evaluation of the integral for  $n=0$ is straightforward, however, 
for $n=1,2$ it requires some tricks which are described e.g. in \cite{resibua}. 

For $n=1$ the basic integrals may be written as 
\begin{eqnarray}
\label{Jn1eval}
J_{k,l,m,1,p,\alpha}=
\int d \vec{g} \int d \vec{C} 
\phi(g) \phi(C) \times 
\nonumber \\
C^k g^l (\vec{C} \cdot \vec{g} )^m \left(\vec{C} \cdot \vec{I}(g) \right)
\end{eqnarray}
with $\vec{g} \equiv \vec{c}_{12}$ and with the vectorial integral
\begin{equation}
\label{I(g)}
\vec{I}(g)\equiv \int d \mu  \, \vec{e}\left|\vec{g} \cdot \vec{e}\right|^{\alpha}
\left(\vec{g} \cdot \vec{e}\right)^p
\end{equation} 
with the short-hand notation 
$d \mu =d\vec{e}\, \Theta(-\vec{g}\cdot \vec{e})\left|\vec{g} \cdot \vec{e}\right|$\,.

Similarly, for $n=2$ one can write
\begin{eqnarray}
\label{Jn1eval1}
J_{k,l,m,2,p,\alpha}=
\int d \vec{g} \int d \vec{C} 
\phi(g) \phi(C) \times 
\nonumber \\
C^k g^l (\vec{C} \cdot \vec{g} )^m \vec{C} \cdot \hat{H}(g) \cdot \vec{C}
\end{eqnarray}
where the dyad $\hat{H}(g)$ is given by
\begin{equation}
\label{H(g)}
\hat{H}(g)\equiv \int d \mu  \, \vec{e} \circ \vec{e}
\left|\vec{g} \cdot \vec{e}\right|^{\alpha}
\left(\vec{g} \cdot \vec{e}\right)^p\,, 
\end{equation}
and where $\circ $ denotes direct vector product. 
Due to symmetry one can write $\vec{I}(g)=\vec{g}G(g)$, where the 
function $G(g)$ may be found from the equation
\begin{equation}
\label{I(g)1}
\vec{g} \cdot \vec{I}(g)= g^2G(g)=
\int d\vec{e} \, \Theta(-\vec{g}\cdot \vec{e})\left|\vec{g} \cdot \vec{e}\right|^{1+\alpha}
\left(\vec{g} \cdot \vec{e}\right)^{p+1}\,.
\end{equation}
The integral in the right-hand side of 
(\ref{I(g)1}) may be evaluated using spherical coordinates:
\begin{equation}
\label{spherint}
\int d\vec{e} \, \Theta(-\vec{g}\cdot \vec{e})\left|\vec{g} \cdot \vec{e}\right|^{\beta}
\left(\vec{g} \cdot \vec{e}\right)^{r}=
\frac{2 \pi (-1)^{r}}{r+\beta+1}g^{r+\beta}\,.
\end{equation}
This yields the function $G(g)$ and, thus, the vectorial integral
\begin{equation}
\label{I(g)2}
\vec{I}(g)= 2\pi (-1)^{p+1}\frac{g^{p+\alpha}}{p+\alpha+3} \, \vec{g}\,.
\end{equation}
For the dyad $\hat{H}(g)$ one can also use symmetry arguments to write
\begin{equation}
\label{H(g)1}
\hat{H}(g)=A(g)\, \vec{g} \circ \vec{g} +B(g)g^2\hat{U}\,,
\end{equation}
where $\hat{U}$ is a unit dyad (i.e. diagonal matrix). Multiplying 
$\hat{H}$ from both sides by $\vec{g}$ and then taking the trace, we
obtain the set of equations for the functions $A(g)$ and $B(g)$:

\begin{eqnarray}
\label{H(g)2}
&&\vec{g} \cdot \hat{H} \cdot \vec{g}= 
Ag^4+Bg^4= \\
&&\int d\vec{e} \, \Theta(-\vec{g}\cdot \vec{e})\left|\vec{g} \cdot \vec{e}\right|^{1+\alpha}
\left(\vec{g} \cdot \vec{e}\right)^{p+2}=
\frac{2 \pi (-1)^{p}}{p+\alpha+4} \, g^{p+\alpha+3} \nonumber 
\end{eqnarray}
and 
\begin{eqnarray}
\label{H(g)3}
&&{\bf Tr} \hat{H} = 
Ag^2+3Bg^2= \\
&&\int d\vec{e} \, \Theta(-\vec{g}\cdot \vec{e})\left|\vec{g} \cdot \vec{e}\right|^{1+\alpha}
\left(\vec{g} \cdot \vec{e}\right)^{p}=
\frac{2 \pi (-1)^{p}}{p+\alpha+2} \, g^{p+\alpha+1} \nonumber 
\end{eqnarray}
Solving the set (\ref{H(g)2}), (\ref{H(g)3}) for $A(g)$ and $B(g)$ we obtain
\begin{equation}
\label{H(g)4}
\hat{H} = 
\frac{2 \pi (-1)^{p}g^{p+\alpha-1}}{(p+\alpha+4)(p+\alpha+2)}
\left[(p+\alpha+1)\,\vec{g} \circ \vec{g}+g^2 \hat{U} \right]\,.
\end{equation}
With Eqs. (\ref{I(g)2}) and (\ref{H(g)4}) the basic integrals 
$J_{k,l,m,n,p,\alpha}$ for $n=1$ and $n=2$ can be reduced to the integrals
\begin{equation}
\label{Jn1n2gen}
\int d \vec{g} \int d \vec{C} 
\phi(g) \phi(C) 
C^{k+\nu_1} g^{l+p+\alpha+\nu_2} 
(\vec{C} \cdot \vec{g} )^{m+\nu_3}
\end{equation}
with $\nu_1=0$, $\nu_2=0$, $\nu_3=1$ to evaluate the integral for $n=1$ and 
with $\nu_1=0$, $\nu_2=-1$, $\nu_3=2$ and 
$\nu_1=2$, $\nu_2=1$, $\nu_3=0$ for $n=2$. 
The computation of these integrals is straightforward and yield the 
final result (\ref{Jn0}), (\ref{Jn1}) and (\ref{Jn2}) which has been 
given above.

\end{multicols}
\end{document}